\documentclass{aa}
\usepackage{natbib}
\bibpunct{(}{)}{;}{a}{}{,}
\usepackage[pdfpagelabels=false]{hyperref}	% Hyperlinks
\hypersetup{colorlinks=true,linkcolor=black,citecolor=black,filecolor=black,urlcolor=black,}
\usepackage{graphicx}
\usepackage{txfonts}
\usepackage{amsmath}	
\usepackage{subcaption}
\usepackage{booktabs}
\usepackage{multirow}
\usepackage{stackengine}
\usepackage{xurl}
\usepackage{caption}
\usepackage{longtable}

\usepackage{silence}
\begin{document}

   \title{Discriminating among cosmological models by data-driven methods}
   \titlerunning{Discriminating cosmology by data-driven methods}

   % \subtitle{}

   \author{S. Vilardi
          \inst{1}
          \and
          S. Capozziello
          \inst{1,2,3} % \orcidlink{0000-0003-4886-2024}
          \and
          M. Brescia\inst{1,3,4} % \orcidlink{0000-0001-9506-5680}
          }

   \institute{Dipartimento di Fisica "E. Pancini", Università degli Studi di Napoli "Federico II", Complesso Univ. Monte S. Angelo, Via             Cinthia 9, I-80126 Napoli, Italy
         \and Scuola Superiore Meridionale, Largo S. Marcellino 10, I-80138 Napoli, Italy
         \and Istituto Nazionale di Fisica Nucleare (INFN), Sez. di Napoli, Complesso Univ. Monte S. Angelo, Via Cinthia 9, I-80126 Napoli, Italy\\
         \begingroup
         \hypersetup{urlcolor=black}
         \email{\href{mailto:capozziello@na.infn.it}{capozziello@na.infn.it}}
         \endgroup
         \and
              INAF – Astronomical Observatory of Capodimonte, via Moiariello 16, I-80131 Napoli, Italy\\
             }

   \date{Received ; accepted }

% \abstract{}{}{}{}{} 
% 5 {} token are mandatory
 
  \abstract
  % context heading (optional)
  % {} leave it empty if necessary  
   {The study examines the Pantheon+SH0ES dataset using the standard Lambda Cold Dark Matter ($\Lambda$CDM) model as a prior and applies machine learning to assess potential deviations. Rather than assuming discrepancies, we test the models’ goodness of fit and explore whether the data allow alternative cosmological features.}
  % aims heading (mandatory)
   {The central goal is to evaluate the robustness of the $\Lambda$CDM model compared with other dark energy models, and to investigate whether there are deviations that might indicate new cosmological insights. The study  takes into account a data-driven approach, using both traditional statistical methods and machine learning techniques.}
  % methods heading (mandatory)
   {Initially, we evaluate six dark energy models using traditional statistical methods like Monte Carlo Markov chain (MCMC) and Static/Dynamic Nested Sampling to infer cosmological parameters. We then adopt a machine learning approach, developing a regression model to compute the distance modulus for each supernova, expanding the feature set to 74 statistical features. This approach uses an ensemble of four models: MultiLayer Perceptron, k-Nearest Neighbours, Random Forest Regressor, and Gradient Boosting. Cosmological parameters are estimated in four scenarios using MCMC and Nested Sampling, while feature selection techniques (Random Forest, Boruta, SHapley Additive exPlanation (SHAP)) are applied in three.}
  % results heading (mandatory)
   {Traditional statistical analysis confirms that the $\Lambda$CDM model is robust, yielding expected parameter values. Other models show deviations, with the Generalised and Modified Chaplygin Gas models performing poorly. In the machine learning analysis, feature selection techniques, particularly Boruta, significantly improve model performance. In particular, models initially considered weak (Generalised/Modified Chaplygin Gas) show significant improvement after feature selection.}
  % conclusions heading (optional), leave it empty if necessary 
   {The study demonstrates the effectiveness of a data-driven approach to cosmological model evaluation. The $\Lambda$CDM model remains robust, while machine learning techniques, in particular feature selection, reveal potential improvements in alternative models which could be relevant for new observational campaigns like the recent Dark Energy Spectroscopic Instrument (DESI) survey.}


   \maketitle
%
%%%%%%%%%%%%%%%%%% BODY OF PAPER %%%%%%%%%%%%%%%%%%

\section{Introduction}
At the turn of the 21st century, a significant breakthrough in our comprehension of the cosmos occurred, thanks to two separate teams of cosmologists. The Supernova Cosmology Project  \citep{perlmutter1999measurements}, and the High-Z Supernovae Search Team  \citep{schmidt1998high} inaugurated a new era of cosmic understanding. By studying far-off Type Ia Supernovae, they uncovered a universe that was expanding at an accelerated pace. The groundbreaking discovery necessitated a re-evaluation of longstanding cosmic assumptions and triggered the development of new models. Most of them are related to the concept of dark energy \citep{Peebles:2002gy}, a cosmic enigma hypothesised to account for the observed accelerated expansion (see \cite{Bamba:2012cp} for a review). Einstein's General Relativity, when applied to cosmology sourced by baryonic  matter and radiation, cannot explain such an accelerated dynamics. As a result, new hypotheses regarding dark energy \citep{Basilakos:2008ae} or extensions of General Relativity \citep{Capozziello:2011et} have become imperative. The $\Lambda$CDM model, which revisits Einstein's original concept of a cosmological constant, has emerged as a strong contender for explaining accelerated dynamics \citep{ostriker1986generation}. This model has revived the repulsive gravitational impact of the cosmological constant as a credible means of explaining cosmic acceleration. In terms of particle physics, the cosmological constant $\Lambda$ is representative of vacuum energy  \citep{Weinberg:1988cp}. Thus, the quest for a mechanism yielding a small, observationally consistent value for the cosmological constant remains paramount. Distinguishing among the myriad dark energy models necessitates the establishment of observational constraints, often derived from phenomena such as Type Ia Supernovae, Cosmic Microwave Background (CMB) radiation, and large-scale structure observations. A key objective to deal with dark energy is  identifying  any potential deviations in the value of the parameter
\begin{equation}
	w_{\text{de}} = \frac{p_{\text{de}}}{\rho_{\text{de}}},
\end{equation}
where $p_{\text{de}}$ is the pressure of dark energy and $\rho_{\text{de}}$ is its energy density, from its standard value of \begin{equation}w_{\Lambda} = -1,\end{equation}and to determine whether it aligns with the cosmological constant or diverges at some cosmic scale. Following the recent Dark Energy Spectroscopic Instrument (DESI) results, a number of studies have emerged that explore the evolving nature of dark energy. In particular, research has suggested that dark energy may not be a constant force, as originally thought, and may be evolving over time \citep{Tada2024, Lili2024}.

The present study is devoted to this issue. The aim is to show that concurring dark energy cosmological models can be automatically discriminated applying machine learning techniques to suitable samples of data.

The paper consists of two interconnected yet distinct parts. The initial section focuses on Bayesian inference using sampling techniques, specifically Monte Carlo Markov Chains (MCMC) \citep{geyer1992practical} and Nested Sampling \citep{skilling2004nested}, applied to the original data set. 

MCMC is a versatile method that can be used to sample from any probability distribution. Its primary use is for sampling from hard-to-handle posterior distributions in Bayesian inference.
In Bayesian estimation, computing the marginalized probability can be computationally expensive, especially for continuous distributions. The key advantage of MCMC is its ability to bypass the calculation of the normalisation constant.
The general idea of the algorithm involves initiating a Markov chain with a random probability distribution over states. It gradually converges towards the desired probability distribution. The algorithm relies on a condition (Detailed Balance Sheet) to ensure that the stationary distribution of the Markov chain approximates the posterior probability distribution.

If this condition is satisfied, it guarantees that the stationary state of the Markov chain approximates the posterior distribution. Although MCMC is a complex method, it offers great flexibility, allowing efficient sampling in high-dimensional spaces and solving problems with large state spaces. However, it has a limitation: MCMC is poor at approximating probability distributions with multiple modes.

Nested Sampling (NS) is a computational algorithm used to estimate evidence (a measure of how well a model fits the data) and infer posterior distributions in Bayesian analysis. Introduced by \citet{skilling2004nested}, it is characterised by its ability to deal efficiently with high-dimensional parameter spaces, such as those found in cosmological studies. Rather than uniformly exploring the parameter space like MCMC, Nested Sampling strategically selects points, called 'live points', which are progressively refined to focus on regions of higher likelihood, making it particularly effective for testing complex cosmological models such as the Generalised and Modified Chaplygin Gas.

Nested Sampling is particularly useful in cosmological analyses where multimodal likelihoods, such as those arising in dark energy model tests, pose a challenge to traditional MCMC. By focusing on regions of high likelihood, Nested Sampling efficiently narrows the plausible parameter space, making it ideal for our study, where we are investigating competing dark energy models that could yield complex, multimodal posterior distributions. This method allows us to compare the likelihood of each model while providing robust parameter estimates, particularly for $\Omega_m$ and $w$, which directly influence our conclusions about the evolution of dark energy. Furthermore, it has a built-in self-tuning capability, allowing immediate application to new problems.

In our work, we are going to use two versions of NS: one with a fixed number of live points, called Static Nested Sampling (SNS), and one with a varying number of live points during runtime, called Dynamic Nested Sampling (DNS). 

Our analysis considers six dark energy parameterisations, each with different properties and free parameters. Rather than assuming a deviation from $\Lambda$CDM, we use a machine learning approach to compare these models and determine which best describes the data. The models considered are:
\begin{enumerate}
    \item $\Lambda$CDM: the standard model.
    \item Linear Redshift \citep{huterer2001probing, weller2002future}: that propose a linear relation between redshift and \textit{w}. It is the simplest possible parameterisation.
    \item Chevallier-Polarski-Linder \citep{chevallier2001accelerating, linder2008dynamics} (CPL): simple, but flexible and robust parameterisation that tries to cover the over-all time evolution of \textit{w}.
    \item Squared Redshift \citep{barboza2008parametric}: this model propose a squared relation between redshift and \textit{w} and covers the universe redshift regions where the CPL parameterisation fails.
    \item Generalised Chaplygin Gas \citep{bento2002generalized} (GCG): is the first scenario that we will investigate where dark matter and dark energy are unified.
    \item Modified Chaplygin Gas \citep{benaoum2012modified} (MCG): is a modified version of the previous parameterisation and has the largest number of free parameters among the models we studied, three.
\end{enumerate}
This scheme allows us to assess how well each model fits the observational data, providing insights into possible variations in the behaviour of dark energy without assuming the need for a new paradigm.

In the second section of this study, inspired by the work of \citet{d2016analysis} and the Feature Analysis for Time Series (\texttt{FATS}) public Python library \citep{nun2015fats}, we are going to compute additional statistics for each supernova and to use three feature selection techniques to identify significant parameters from a final set of $70$ features.
In fact, as we will see, it is possible to  analyse four different cases: a 'base' case where no feature selection is used; a case where the first 18 features selected by Random Forest are taken \citep{liaw2002classification}; a case where the feature selection method used is Boruta \citep{kursa2011all}; and a last case where the first 18 features selected by SHAP are taken into account \citep{lundberg2017unified}.
An ensemble learning strategy is then utilised to create a predictive model for the distance modulus based on the selected features. The models we are going to use in the ensemble learning are the following:
\begin{enumerate}
    \item MultiLayer Perceptron \citep{rumelhart1986learning} (MLP): a modern feedforward artificial neural network, consisting of fully connected neurons with a non-linear kind of activation function.
    \item k-Nearest Neighbours \citep{cover1967nearest} (k-NN): a non-parametric supervised learning method used for both classification and regression. 
    \item Random Forest Regressor \citep{breiman2001random}: an ensemble learning method for classification, regression and other tasks that operates by constructing a multitude of decision trees at training time. For regression tasks, the mean or average prediction of the individual trees is returned.
    \item Gradient Boosting \citep{schapire1990strength}:  a machine learning technique used in regression and classification tasks, among others. It gives a prediction model in the form of an ensemble of weak prediction models, that is, models  making very few assumptions about the data, which are typically simple decision trees \citep{quinlan1986induction}. When a decision tree is the weak learner, the resulting algorithm is called gradient-boosted trees \citep{friedman2001greedy}.
\end{enumerate}
Subsequently, the new dataset, composed by original redshifts and predicted distance moduli, is subjected to the same sampling techniques previously used in the derivation of cosmological parameters.

The paper is structured as follows: Sect. \ref{sec:DE} outlines the six dark energy models analysed. In Sect. \ref{sec:DATA}, the data set is presented. We  describe the compilation of Pantheon+SH0ES and the new features that have been added. Sect. \ref{sec:ML} focuses on feature selection techniques used and the models implemented in our ensemble learning. In Sect. \ref{sec:MCMC}, the different sampling techniques  are described in detail, with an explanation of the specifics of MCMC and NS for the inference of cosmological parameters. We conclude with a brief introduction to the information criteria used to evaluate the performance of the techniques. Sect. \ref{sec:RES} presents the results of the study, showing the insights gained by implementing traditional Bayesian inference techniques as well as machine learning and sampling methods. Finally, in Sect. \ref{sec:CONCL}, we summarise our conclusions and present some perspectives of the approach.
%--------------------------------------------------------------------
\section{Dark energy models}
\label{sec:DE}
%-------------------------------------- 
 There are several alternatives proposed to the $\Lambda$CDM model, ranging from adding phenomenological dark energy terms to modifying the Hilbert-Einstein action or considering other geometrical invariants \citep{Cai:2015emx}. To refine the model with dark energy evolving over time, a barotropic factor $\omega(z) = P/\rho$ dependent on $z$ can be considered. This is the equation of state (EoS) of the given cosmological model. However, this approach has a crucial aspect: it is not possible to define $\omega(z)$ \(\textit{a priori}\); it must be reconstructed starting from observations. As stated in \citet{dunsby2016theory}, it is advantageous to express the barotropic factor in terms of cosmic time or, more appropriately, as a function of the scale factor or redshift. This choice is based on the idea that dark energy could evolve through a generic function over the history of the universe. In this sense, the cosmographic analysis can greatly help in reconstructing the cosmic flow by the choice of suitable polynomials in the redshift $z$ (see e.g. \cite{Demianski:2012ra, Capozziello:2017nbu, Capozziello:2020ctn, Capozziello:2021xjw, Benetti:2019gmo}). A straightforward approach involves expanding \(\omega\) as a Taylor series in redshift \(z\):

\begin{equation}
    \omega(z) = \sum_{n=0}^{\infty} \omega_n z^n.
\end{equation}

However, opting for this expansion could pose challenges, as it may lead to a divergence in the equation of state at higher redshifts.

While certain models, such as the Linear Redshift and Chaplygin Gas, have been challenged by past studies \citep{Fabris2011}, we deliberately include a diverse set of parameterizations. Our goal is to use machine learning techniques to evaluate their relative performance across the dataset, rather than presupposing their validity or exclusion. This approach allows for an unbiased assessment of different dark energy descriptions, including both commonly accepted and alternative models.

The following paragraphs present the six models studied in this paper.
\subsection{\texorpdfstring{$\Lambda$}{Lambda}CDM}
The $\Lambda$CDM model is the standard cosmological model, characterised by $w = -1$, with the Hubble function given by:
\begin{equation}
E(z)^2 = \left(\frac{H(z)}{H_0}\right)^2 = \omega_m (1 + z)^3 + (1 - \omega_m).
\end{equation}
While $\Lambda$CDM provides an excellent fit to a wide range of observational data, it does not address certain fundamental issues, such as the cosmic coincidence problem or the fine-tuning of $\Lambda$.
\subsection{Linear redshift parameterisation}
The linear redshift model is one of the simplest extensions of the $\Lambda$CDM, introducing a redshift-dependent equation of state (EoS) for dark energy:
\begin{equation}
    w(z) = w_0 - w_a z,
\end{equation}
where $w_0$ and $w_a$ (sometimes written as $w_z$) are constants, with $w_0$ representing the present value of $ w(z) $. The model reduces to $\Lambda$CDM for $w_0 = -1$ and $w_a = 0$. The corresponding Hubble function is
\begin{equation}
    E(z)^2 = \Omega_m (1 + z)^3 + \Omega_x (1 + z)^{3(1+w_0+w_a)}e^{-3w_a z},
\end{equation}
where $\Omega_m$ is the matter density parameter and $\Omega_x$ is the dark energy density. However, this parameterisation diverges at high redshifts, requiring strong constraints on $w_a$ in studies using high redshift data, such as CMB observations \citep{wang2006constraining}.
\subsection{Chevallier-Polarski-Linder (CPL) Parameterisation}
The CPL parameterisation introduces a smoothly varying EoS with two parameters characterising the present value ($w_0$) and its evolution with time:
\begin{equation}
    w(z) = w_0 + \frac{z}{1+z} w_a.
\end{equation}
The corresponding Hubble function is given by \citep{escamilla2019unveiling}:
\begin{equation}
    E(z)^2 = \Omega_m (1 + z)^3 + \Omega_x (1 + z)^{3(1+w_0+w_a)} e^{-\frac{3w_a z}{1+z}}.
\end{equation}
The CPL model is widely used because of its flexibility and robust behaviour in describing the evolution of dark energy.
\subsection{Squared Redshift Parameterisation}
This model provides an improvement over the CPL in regions where the latter cannot be reliably extended to describe the entire cosmic history. Its functional form is
\begin{equation}
    w(z) = w_0 + \frac{z(1 + z)}{1 + z^2} w_a,
\end{equation}
which remains well behaved as $z \to -1$. The corresponding Hubble function is
\begin{equation}
    E(z)^2 = \Omega_m(1 + z)^3 + (1 - \Omega_m)(1 + z)^{3(1+w_0)}(1 + z^2)^{\frac{3w_a}{2}}.
\end{equation}
Overall, the squared redshift parameterisation has the advantage of remaining finite throughout the history of the Universe.
\subsection{Unified Dark Energy Fluid scenarios}
As an extension of conventional cosmological scenarios, within the framework of a homogeneous and isotropic universe, we assume that the gravitational sector follows the standard formulation of General Relativity, with minimal coupling to the matter sector. Furthermore, we assume that the total energy content of the universe consists of photons ($\gamma$), baryons ($b$), neutrinos ($\nu$) and a unified dark fluid (UDF, $X_U$) \citep{Cardone:2004sq,paul2013observational}. This UDF is capable of exhibiting properties characteristic of dark energy, dark matter or an alternative cosmic fluid as the universe expands. Consequently, the total energy density is denoted as $\rho_i$, where $i = \gamma, b, \nu, X_U$. Each fluid component obeys a continuity equation of the form
\begin{equation}
    \dot{\rho_i} + 3\frac{\dot{a}}{a}(\rho_i + p_i) = 0.
\end{equation}
Standard solutions give $\rho_b \propto a^{-3}$ and $\rho_{\gamma, \nu} \propto a^{-4}$. For the UDF we assume a constant adiabatic sound velocity $c_s$ and express the pressure as $p = c_s^2 (\rho - \tilde{\rho})$, where $c_s$ and $\tilde{\rho}$ are positive constants \citep{escamilla2020deep}. This formulation allows the fluid to exhibit both barotropic and $\Lambda$-like behaviour, effectively unifying dark matter and dark energy, a phenomenon known as dark degeneracy.
By integrating the continuity equation for the UDF, we obtain
\begin{equation}
    \rho = \rho_{\Lambda} + \rho_{X_U} a^{-3(1+c_s^2)},
\end{equation}
\begin{equation}
    p = -\rho_{\Lambda} + c_s^2 \rho_{X_U} a^{-3(1+c_s^2)},
\end{equation}
where $\rho_{\Lambda} = \frac{c_s^2 \tilde{\rho}}{1+c_s^2}$ and $\rho_{X_U} = \rho_0 - \rho_{\Lambda}$, where $\rho_0$ is the present dark energy density. The dynamical equation of state (EoS) is given by
\begin{equation}
    w = -1 + \frac{1 + c_s^2}{\left(\frac{\rho_{\Lambda}}{\rho_{X_U}}\right)(1 + z)^{-3(1+c_s^2)} + 1}.
\end{equation}
To fully describe the behaviour of $X_U$, we need to specify a particular functional form for $p_{X_U}$ in terms of $\rho_{X_U}$.
\subsubsection{Generalised Chaplygin Gas Model}
The Generalised Chaplygin Gas (GCG) model characterises $X_U$ by the equation of state:
\begin{equation}
     p_{\text{gcg}} = -\frac{A}{(\rho_{\text{gcg}})^{\alpha}},
\end{equation}
where $A$ and $0 \leq \alpha \leq 1$ are free parameters. The case $\alpha = 1$ corresponds to the original Chaplygin gas model. Solving the continuity equation gives the evolution of the energy density:
\begin{equation}
    \rho_{\text{gcg}}(a) = \rho_{\text{gcg},0} \left[b + (1 - b)a^{-3(1+\alpha)}\right]^{\frac{1}{1+\alpha}},
\end{equation}
where $\rho_{\text{gcg},0}$ is the current energy density and $b = A \rho_{\text{gcg},0}^{-(1+\alpha)}$. The corresponding dynamical equation of state is
\begin{equation}
    w_{\text{gcg}}(z) = -\frac{b}{b + (1 - b) (1+z)^{-3(1+\alpha)}}.
\end{equation}
This model describes an effective transition between dark matter and dark energy behaviour, with an intermediate regime when $\alpha = 1$.
\subsubsection{Modified Chaplygin Gas Model}
The Modified Chaplygin Gas (MCG) model extends the GCG by introducing a linear term in the pressure-density relation:
\begin{equation}
    p_{\text{mcg}} = b\rho_{\text{mcg}} - \frac{A}{(\rho_{\text{mcg}})^{\alpha}},
\end{equation}
where $A$, $b$ and $\alpha$ are real constants with $0 \leq \alpha \leq 1$. Setting $A = 0$ yields a perfect fluid with $w = b$, while $b = 0$ restores the GCG model. The standard Chaplygin gas model corresponds to $\alpha = 0$. The evolution of the energy density follows:
\begin{equation}
    \rho_{\text{mcg}}(a) = \rho_{\text{mcg},0} \left[b_s + (1 - b_s)a^{-3(1+b)(1+\alpha)}\right]^{\frac{1}{1+\alpha}},
\end{equation}
where $\rho_{\text{mcg},0}$ is the current MCG energy density, and $b_s = A \rho_{\text{gcg},0}^{-(1+\alpha)}/(1 + b)$. The corresponding equation of state becomes
\begin{equation}
    w_{\text{mcg}}(z) = b - \frac{b_s(1 + b)}{b_s + (1 - b_s) (1+z)^{-3(1+b)(1+\alpha)}}.
\end{equation}
As an extension of the GCG model, the MCG retains similar behaviour across different cosmological epochs. The Chaplygin gas framework provides a versatile approach to studying the interplay between dark matter and dark energy throughout cosmic history \citep{Yang:2019nhz}. The interactions between these components can further elucidate the expansion dynamics of the Universe \citep{Piedipalumbo:2023dzg}.

\section{Data Set}
\label{sec:DATA}

As mentioned, the used dataset  is the Pantheon+SH0ES of 1701 Type Ia Supernovae coming from a compilation of 18 different surveys covering a redshift range up to 2.26. Among the 1701 objects in the dataset, 151 are duplicates, observed in multiple surveys, and 12 are pairs or triplets of (Supernova) SN siblings, SNe found in the same host galaxy.

The number of features provided by the Pantheon+SH0ES dataset is 45, excluding the ID of the supernova, the ID of the survey used for that observation, and a binary variable to distinguish the SNe used in SH0ES from those not included. However, to increase the reliability of our model predictions and better capture the intrinsic variability of Type Ia Supernovae, we expanded the feature set from the original 45 features provided by the Pantheon+SH0ES dataset to 71 (still excluding the previously cited features) by incorporating additional statistical descriptors from \citet{d2016analysis}  and the \texttt{FATS} Python library. These additional features help to account for observational uncertainties and intrinsic scatter in the supernova measurements, improving our ability to discriminate between cosmological models, especially in scenarios where small variations in the distance modulus could be critical. For more information on this statistical parameter space, see the Appendix.

\subsection{The Pantheon+SH0ES compilation}

The Pantheon+SH0ES dataset consists of 1701 Type Ia Supernovae (SNeIa) from 18 different surveys, spanning a redshift range from 0.001 to 2.26. This wide range provides valuable insights into the evolution of dark energy over cosmic time. The detailed distance moduli of each supernova in the dataset serve as critical measurements for constraining key cosmological parameters, such as the Hubble constant (H$_0$), the matter density ($\Omega_m$), and the dark energy equation of state parameter ($w$). This comprehensive data set is particularly useful for testing different dark energy models and assessing their consistency with the standard $\Lambda$CDM model.

The theoretical distance modulus ($\mu$) is related to the luminosity distance ($d_L$) by the equation:

\begin{equation} \mu(z) = 5 \log{\frac{d_L(z)}{1 \text{ Mpc}}} + 25, \end{equation}

where $d_L$ is expressed in megaparsecs (Mpc). To account for systematic uncertainties, standard analyses include a nuisance parameter $M$, which represents the unknown offset corresponding to the absolute magnitude of the supernovae and is degenerate with the value of $H_0$.

Assuming a flat cosmological model, the luminosity distance is related to the comoving distance ($D$) by:

\begin{equation} d_L(z) = \frac{c}{H_0} (1 + z) D(z), \end{equation}

where $c$ is the speed of light. The normalised Hubble function ($H(z)/H_0$) is then computed by taking the inverse derivative of $D(z)$ with respect to redshift:

\begin{equation} D(z) = \frac{H_0}{c} \int_0^z \frac{d\tilde{z}}{H(\tilde{z})}. \end{equation}

Here $H_0$ is assumed to be a prior value for normalising $D(z)$.

\section{Methods}
\label{sec:ML}

In our study, we use three different feature selection techniques to identify significant parameters from a final set of 70 features. Our analysis includes four different cases: a baseline scenario with no feature selection, a scenario using the first 18 features selected by Random Forest, a scenario using Boruta feature selection, and a scenario using the first 18 features selected by SHAP. We chose these methods because they provide interpretability in the feature selection process and are well suited to handling the non-linear relationships expected in supernova data. Random Forest and Boruta identify feature importance based on decision tree splits, while SHAP values provide a game-theoretic measure of each feature's contribution to the model's predictions. Other methods, such as Principal Component Analysis (PCA), were not used because they transform features into linear combinations of the original variables, making it difficult to retain direct physical interpretability. Similarly, Recursive Feature Elimination (RFE) was not used because it selects features based on the performance of a particular model, which can introduce bias and limit generalisability across different learning algorithms.

We then use an ensemble learning approach to develop a predictive model for the distance modulus based on the selected features. The ensemble consists of four models: MultiLayer Perceptron (MLP), k-Nearest Neighbours (k-NN), Random Forest Regressor and Gradient Boosting. Each of these models brings unique capabilities to the ensemble, from the flexible architecture of MLP to the non-parametric nature of k-NN, the ensemble learning of Random Forest, and the gradient-boosted trees approach of Gradient Boosting.

We decided to use these models because they strike a balance between flexibility, interpretability and performance in a complex, non-linear problem. Gaussian Processes (GPs) were not used because they do not scale well with large datasets (such as Pantheon+SH0ES) due to their cubic complexity in training. Linear Regression was considered, but is not well suited to capturing non-linear dependencies in the data, making it a poor choice for modelling supernova distance modules. The chosen ensemble approach exploits the strengths of several algorithms to improve robustness and generalisation.

\subsection{Feature selection techniques}

Feature selection is a crucial step in the process of building machine learning models, playing a pivotal role in enhancing model performance, interpretability, and efficiency. In many real-world scenarios, datasets often contain a multitude of features, and not all of them contribute equally to the predictive task at hand. Some features may even introduce noise or lead to computational inefficiencies.

\subsubsection{Random Forest}

In the building of the single Decision Trees, the feature selected at each node is the one which minimises the chosen loss function (like the mean squared error in our case). Feature importance in a Random Forest is calculated based on how much each feature contributes to the reduction in loss function across all the trees in the ensemble. The more frequently a feature is used to split the data and the higher the loss function reduction it achieves, the more important it is considered. 
In Random Forests, 'impurity' refers to the degree of disorder or uncertainty in a decision tree. A 'loss function' quantifies this disorder and measures how well the model is performing. If a feature (such as redshift or luminosity distance) reduces the impurity at multiple decision points within the tree ensemble, it is considered important \citep{liaw2002classification}. This averaged reduction across all trees is used to assess the overall contribution of each feature in predicting cosmological parameters.

\subsubsection{Boruta}

The second method used is an all-relevant feature selection method, or Boruta. The Boruta algorithm takes its name from a demon in Slavic mythology who lived in pine forests and preyed on victims by walking like a shadow among the trees. And, in fact, main concept behind this method is the introduction of \textit{shadow features} and the use of random forest as predicting model \citep{kursa2011all}. A shadow feature for each real one is introduced by randomly shuffling its values among the N samples of the given dataset. It uses a random forest classifier, and so is a feature selection wrapping method, on this extended data set (real and shadow features) and applies a feature importance measure such as Mean Decrease Accuracy and evaluates the importance of each feature. At every iteration, Boruta algorithm checks whether a real feature has a higher importance than the best of its shadow features and constantly removes features which are deemed highly unimportant. Finally, the Boruta algorithm stops either when all features gets confirmed or rejected or it reaches a specified limit of iterations. 

In conclusion, the steps of this algorithm can be summarised like this:
\begin{enumerate}
    \item Take the original features and make a shuffled copy. The new extended dataset is now composed by the original features and their shuffled copy, the shadow features.
    \item Run a random forest classifier on this new dataset and calculate the feature importance of every feature.
    \item Store the highest feature importance of the shadow features and use it as a threshold value.
    \item Keep the original features which have an importance higher than the highest shadow feature importance. We will say that these features make a \textit{hit}.
    \item Repeat the previous steps for some iterations and keep track of the \textit{hits} of the original features.
    \item Label as \textit{confirmed} or \textit{important} the features that have a significantly high number of \textit{hits}; as \textit{rejected} the ones that instead have a significantly low number of \textit{hits}; as \textit{tentative} the ones that fall in between.
\end{enumerate}
The algorithm stops when all features have an established decision, or when a pre-set maximal number of iterations is reached.

\subsubsection{SHAP}

The final feature selection method used in our study was SHAP \citep{lundberg2017unified} (SHapley Additive exPlanations). SHAP adopts a game-theoretic approach to explain the output of machine learning models, connecting optimal credit allocation with local explanations using classic Shapley values from game theory and their related extensions. SHAP serves as a set of software tools designed to enhance the explainability, interpretability, and transparency of predictive models for data scientists and end-users \citep{lundberg2020local}. SHAP is used to explain an existing model. In the context of a binary classification case built with a sklearn model, the process involves training, tuning, and testing the model. Subsequently, SHAP is employed to create an additional model that explains the classification model.
\begin{figure}
    \centering
    \includegraphics[width=\columnwidth]{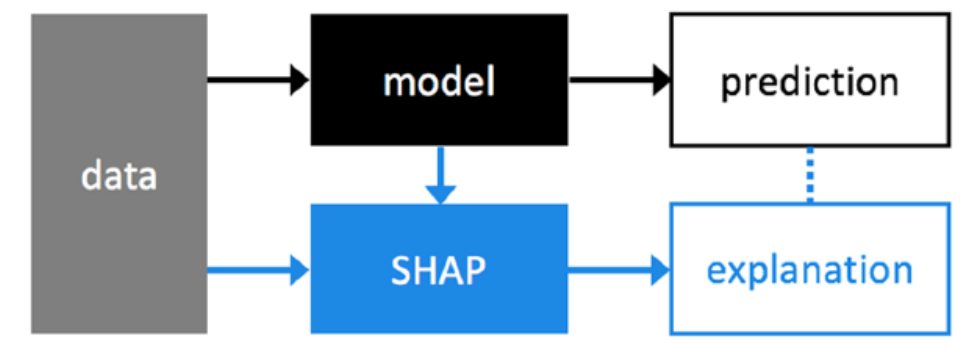}
    \caption[SHAP architecture]{\small SHAP architecture \citep{shaparc}.}
    \label{fig:SHAP architecture}
\end{figure}

The key components of a SHAP explanation include:
\begin{itemize}
    \item[--] explainer: the type of explainability algorithm chosen based on the model used.
    \item[--] base value: it represents the value that would be predicted if no features were known for the current output, typically the mean prediction for the training dataset or the background set. Also called as \textit{reference value}.
    \item[--] SHAPley values: the average contribution of each feature to each prediction for each sample based on all possible features. It is a $(n,m)$ matrix, $n$ samples, $m$ features, that represents the contribution of each feature to each sample.
\end{itemize}

\begin{figure}
    \centering
    \includegraphics[width=\columnwidth]{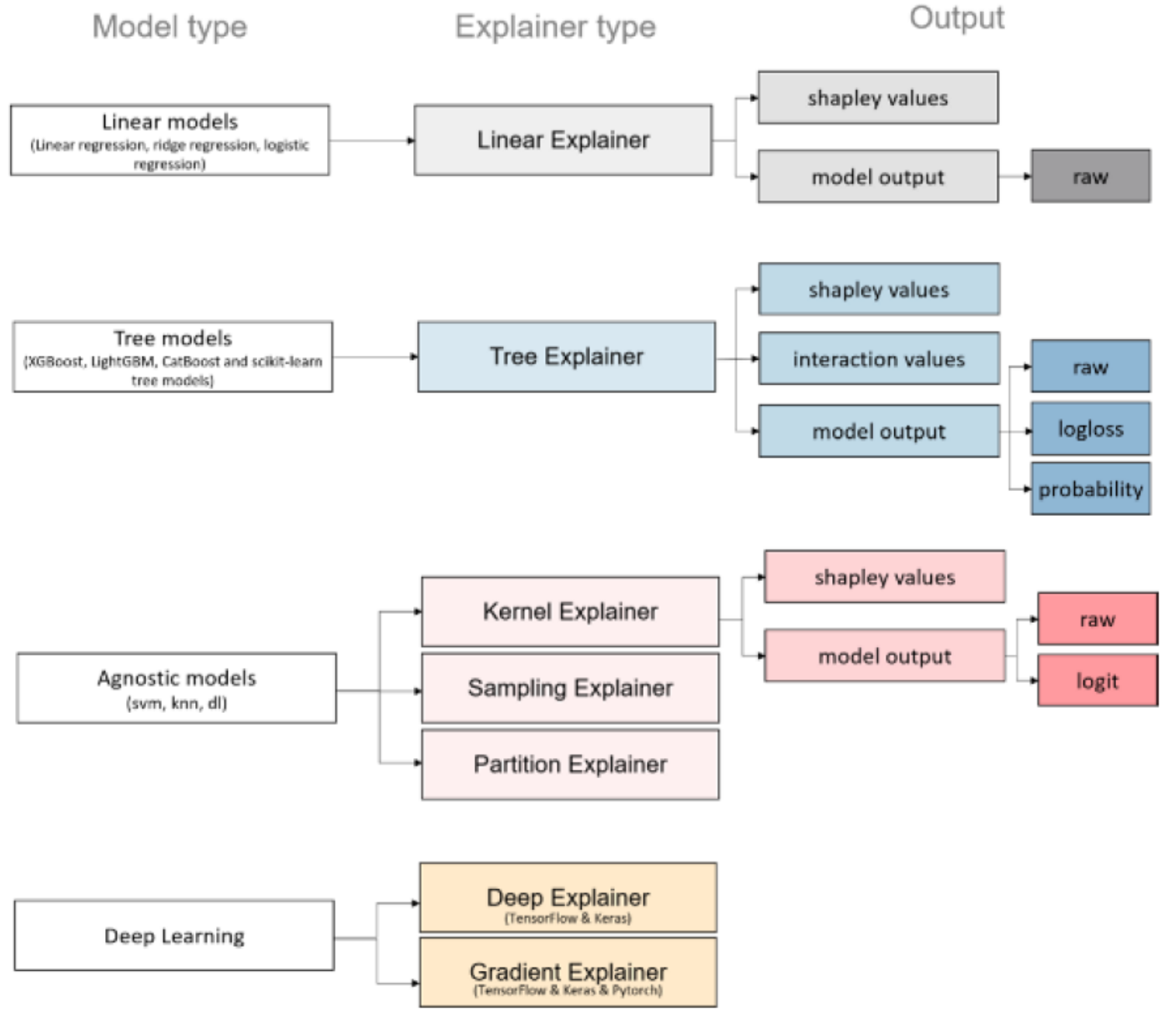}
    \caption[\small Types of SHAP Explainers]{Types of Explainers \citep{class}.}
    \label{fig:Types of Explainers}
\end{figure}
Explainers are the models used to calculate shapley values. The diagram above (Fig. \ref{fig:Types of Explainers}) shows different types of Explainers. The choice of Explainers depends mainly on the selected learning model.
The Kernel Explainer creates a model that substitutes the closest to our model. It also can be used to explain neural networks. For deep learning models, there are the deep and gradient Explainers. In our work we used a Tree Explainer.
Shapley values calculate feature importance by evaluating what a model predicts with and without each feature. Since the order in which a model processes features can influence predictions, this comparison is performed in all possible ways to ensure fair assessments. This approach draws inspiration from game theory, and the resulting Shapley values facilitate the quantification of the impact of interactions between two features on predictions for each sample. As the Shapley values matrix has two dimensions (samples x features), interactions are represented as a tensor with three dimensions (samples x features x features).
\begin{figure}
  \centering
  \begin{subfigure}[t]{0.3\columnwidth}
    \includegraphics[width=\columnwidth]{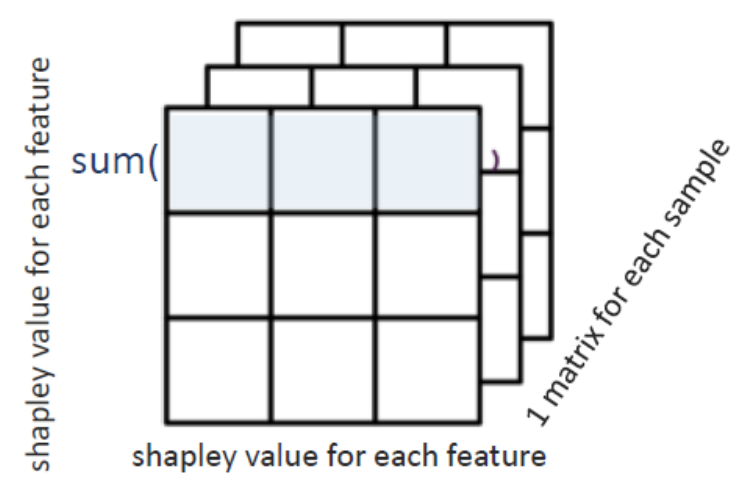}
    \captionsetup{font=scriptsize}
    \caption{Sum of row of matrix (features x features) equals SHAP value of this feature and this sample}
    \label{fig:Sum of row of SHAP matrix}
  \end{subfigure}
  \hfill
  \begin{subfigure}[t]{0.3\columnwidth}
    \includegraphics[width=\columnwidth]{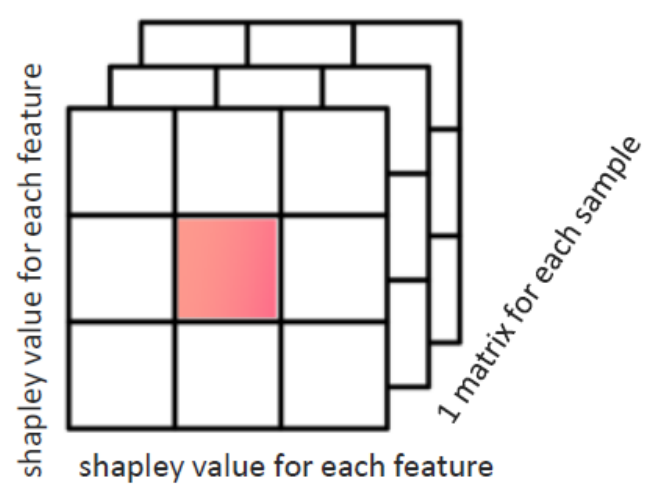}
    \captionsetup{font=scriptsize}
    \caption{The diagonal entries equals the \textit{main effect} of this feature on the prediction}
    \label{fig:Diagonal entries of SHAP matrix}
  \end{subfigure}
  \hfill
  \begin{subfigure}[t]{0.3\columnwidth}
    \includegraphics[width=\columnwidth]{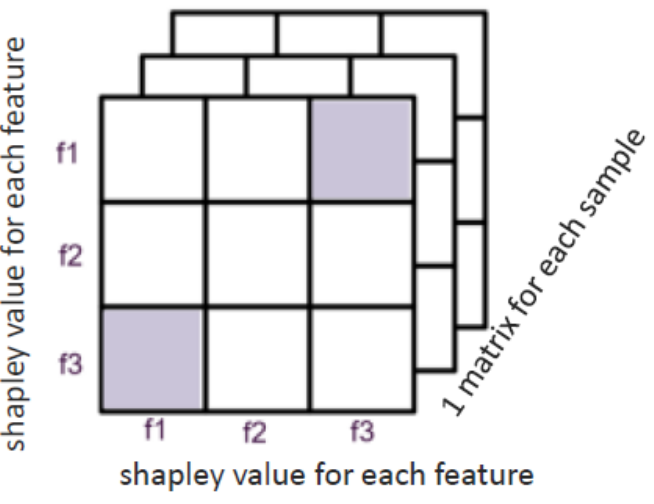}
    \captionsetup{font=scriptsize}
    \caption{The symmetrical entries out of the diagonal equals to the interaction effect between all the pairs of features for a given sample}
    \label{fig:Symmetrical entries of SHAP matrix}
  \end{subfigure}
  \caption[SHAP matrix]{\small SHAP matrix \citep{class}.}
  \label{fig:SHAP matrix}
\end{figure}

\subsection{Ensemble learning}

From the plethora of different machine learning techniques available, the one used in this work is the Ensemble Learning. In this approach, two or more models are fitted on the same data and the predictions from each model are combined. The goal of ensemble learning is to achieve better performance with the ensemble of models than with each individual model by mitigating the weaknesses of each individual model. The models that compose the ensemble learning used in the work will be discussed in the following paragraphs.

\subsubsection{Multi Layer Perceptron}

The MLP, or Multi Layer Perceptron, is the first of the four models used in the ensemble learning used in the work. The MLP is one of the most common used feed forward neural network model \citep{van1986frank, rumelhart1986learning, Brescia2015, Brescia2019} and comes from the profound limitations of the first Rosenblatt's Perceptron in the treatment of non linearly separable, noisy and non numerical data. The term feed-forward refers to the fact that in this neural network model, the impulse is always propagated in the same direction, e.g. from the input layer to the output layer, passing through one or more hidden layers, by combining the sum of weights associated to all neurons except the input ones. The output of each neuron is obtained by an activation function applied to the weighted sum of the inputs. The shape of the activation function can vary considerably from model to model, from the simplest linear function to the hyperbolic tangent, which is the one used in this work. In the training phase of the network, the weights are modified according to the learning rule used, until a predetermined distance between the network output and the desired output is reached (usually this distance is decided \textit{a priori} by the user and is commonly known as the \textit{Error Threshold}).

The easiest way to employ gradient information is to choose the weight update to make small steps in the direction of the negative gradient, so that
\begin{equation}
{w}^{\tau+1} = {w}^{\tau} - \eta\nabla E({w}^{\tau}), 
\end{equation}
where the parameter $\eta > 0$ is referred to as the \textit{learning rate}.  In each iteration, the vector is adjusted in the direction of the steepest decrease of the error function, and this strategy is called \textit{gradient descent}. We still need to define an efficient technique to find the gradient of the error function $E(w)$. A widely used method is the \textit{error backpropagation} in which information is sent alternately forward and backward through the network. However, this method lacks precision and optimisation for complex real-life applications. Therefore, modifications are necessary.

Adaptive Moment Estimation \citep{kingma2014adam} (ADAM) takes a step forward in the pursuit of the minimum of the objective function by solving the problem of \textit{learning rate} selection and avoiding saddle points. ADAM computes adaptive learning rates for each parameter and maintains an exponentially decaying average of past gradients. This average is weighted with respect to the first two statistical moments of the gradient distribution. Adam behaves like a heavy ball with friction, where $\widehat{m_t}$ and $\widehat{v_t}$ are the estimate of the first and second moment of the gradients, and are computed like this:
\begin{equation}
    \widehat{m_t} = \frac{m_t}{1 - \beta_1^t} = \frac{\beta_1 m_{t-1} + (1 - \beta_1)\nabla_wf(W;x^{(i)}y^{(i)})}{1 - \beta_1^t},
\end{equation}
\begin{equation}
    \widehat{v_t} = \frac{v_t}{1 - \beta_2^t} = \frac{\beta_2 v_{t-1} + (1 - \beta_2)\nabla_wf(W;x^{(i)}y^{(i)})^2}{1 - \beta_2^t},
\end{equation}
where $\beta_1$ and $\beta_2$ are the characteristic memory times of the first and second moment of the gradients and control the decay of the moving averages. The final formula is then:
\begin{equation}
    W_{t+1} = W_t - \frac{\eta}{\sqrt{\widehat{v_t} + \epsilon}}\widehat{m_t}.
\end{equation}
In summary, ADAM's advantage lies in its use of the second moment of the gradient distribution.

The hyperparameters used to build our MLP are the following:
\begin{itemize}
    \item[--] two hidden layers with $100$ neurons each;
    \item[--] the \textit{tanh} as activation function;
    \item[--] $1000$ epochs;
    \item[--] the initial learning rate set to 0.01;
    \item[--] ADAM as optimisation technique.
\end{itemize}

\subsubsection{k-Nearest neighbours}

The second model used in our work was the k-Nearest neighbours (k-NN) and it is a non-parametric supervised learning method used for both classification and regression. 

For regression problems, like our work, the k-NN works like this:
\begin{enumerate}
    \item Choose a value for $k$: this determines the number of nearest neighbours used to make the prediction.
    \item Calculate the distance: we calculate the distance between each data point in the training set and the target data point for which a prediction is made.
    \item Find the $k$ nearest neighbours: after calculating the distances, we identify the $k$ nearest neighbours by selecting the $k$ data points nearest to the new data point.
    \item Calculate the prediction: after finding the $k$ neighbours we calculate the value of the dependent variable for the new data point. For this, we take the average of the target values of the $k$ nearest neighbours. Usually, the value of the points are weighted by the inverse of their distance.
\end{enumerate}

For classification problems, a class label is assigned on the basis of a majority vote, i.e. the label that is most frequently represented around a given data point is used.

\begin{figure}
    \centering
    \includegraphics[width=\columnwidth]{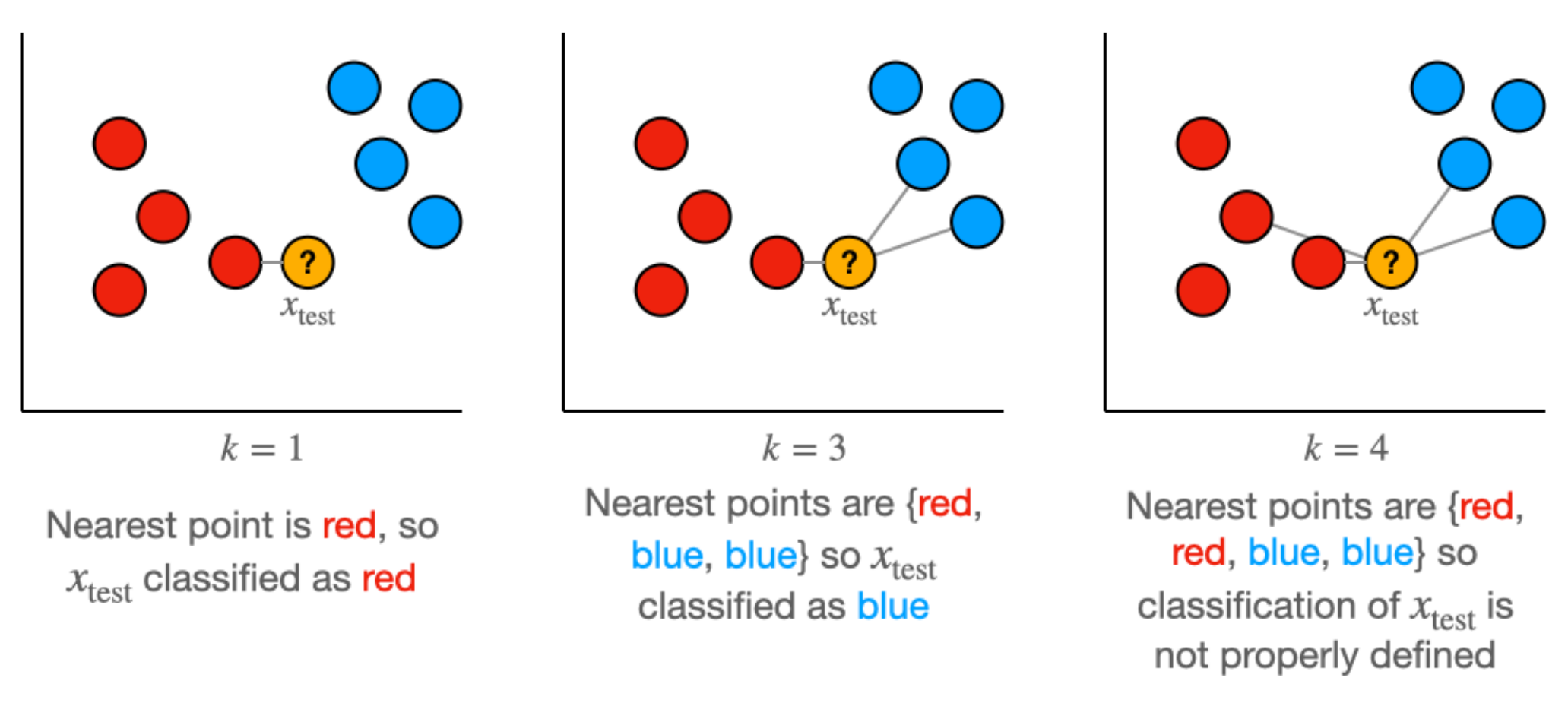}
    \caption[k-NN Classification]{\small k-NN Classification. A simple solution for the last case is to randomly select one of the two classes or use an odd $k$ \citep{knn}.}
    \label{fig:k-NN classification}
\end{figure}

Three different algorithms are available to perform k-NN:
\begin{itemize}
    \item[--] Brute Force: here we simply calculate the distance from the point of interest to all the points in the training set and take the class with majority points. 
    \item[--] k-Dimensional Tree \citep{bentley1975multidimensional} (kd tree): kd tree is a hierarchical binary tree. When this algorithm is used for k-NN classification, it rearranges the whole dataset in a binary tree structure, so that when test data is provided, it would give out the result by traversing through the tree, which takes less time than brute search.
    \item[--] Ball Tree \citep{bhatia2010survey}: is a hierarchical data structure similar to kd trees and is particularly efficient for higher dimensions.
\end{itemize}

The hyperparameters selected for constructing our k-NN model are determined using \textit{GridSearchCV} (Grid Search Cross Validation) from the \textit{sklearn} Python library \citep{scikit-learn}. This method identifies the optimal combination of hyperparameters from a predefined parameter grid based on a specified scoring function, with the negative mean squared error employed in our work. Additionally, GridSearchCV utilizes cross-validation to refine the model parameters, and in our case, a \textit{cv} value of 10 was applied. The ultimate hyperparameters are as follows:
\begin{itemize}
    \item[--] the number of neighbours, the \textit{k} value, is 4 when no feature selection is employed, 5 for feature selection with the Random Forest and Boruta, 6 when the feature selection is done with SHAP;
    \item[--] the weight is the inverse of the distance used;
    \item[--] the power parameter \textit{p} for the Minkowski metric is 1, so we used the Manhattan distance.
    \item[--] the algorithm hyperparameter used to compute the nearest neighbours was leaved to \textit{auto}, so that the model will automatically use the most appropriate algorithm based on the values passed to the \textit{fit} method.
\end{itemize}

\subsubsection{Random Forest}

The third model utilized in our study is the Random Forest Regressor (RFRegressor). This model works by creating an ensemble of Decision Trees during the training phase, each based on different subsets of input data samples. Within the construction of each tree, various combinations of features inherent in data patterns are incorporated into the decision-making process. By employing a sufficient number of trees (dependent on the problem space complexity and input data volume), the produced forest is likely to represent all given features \citep{hastie2009overview}. Regression models, in general sense, are able to take variable inputs and predict an output from a continuous range. In the context of regression models, which predict an output within a continuous range, decision tree regressions typically lack the ability to produce continuous output. Instead, they are trained on examples with output lying in a continuous range. 

The hyperparameters used to build our RFRegressor model were the following:
\begin{itemize}
    \item[--] the number of trees is 10000;
    \item[--] the criterion to measure the quality of a split is the mean squared error; 
    \item[--] the maximum depth of the trees is set to \textit{None}, so the nodes are expanded until all leaves are pure or until all leaves contain less than \textit{min\_samples\_split} samples;
    \item[--] the minimum number of samples required to split an internal node, the hyperparameter of the previous point \textit{min\_samples\_split}, is set to 2;
    \item[--] the number of feature to consider when looking for the best split hyperparameter is set to \textit{auto}, so that the max features to consider is equal to the number of features available.
\end{itemize}

\subsubsection{Gradient Boosting}

The fourth and last model used in our work is the Gradient Boosting Regressor (GBRegressor). In the previous paragraph we talked about the \textit{bagging} technique, here the technique is called \textit{boosting} and is somehow complementary. Boosting is a sequential type of ensemble learning that uses the result of the previous model as input for the next one. Instead of training the models separately, the upgrade trains the models in sequence, each new model being trained to correct the errors of the previous ones. At each iteration the correctly predicted results are given a lower weight and those erroneously predicted a greater weight. It then uses a weighted average to produce a final result. 
Boosting is an iterative meta-algorithm that provides guidelines on how to connect a set of Weak Learners to create a Strong Learner. The key to the success of this paradigm lies in the iterative construction of Strong Learners, where each step involves introducing a Weak Learner tasked with "adjusting the shot" based on the results obtained by its predecessors. Gradient Boosting employs standard Gradient Descent to minimize the loss function used in the process. Typically, the Weak Learners are decision trees, and in this case, the algorithm is termed gradient-boosted trees.

The typical steps of a Gradient Boosting algorithm are the following:
\begin{enumerate}
    \item The average of the target values is calculated for the initial predictions and the corresponding initial residual errors.
    \item A model (shallow decision tree) is trained with independent variables and residual errors as data to obtain predictions.
    \item The additive predictions and residual errors are calculated with a certain learning rate from the previous output predictions obtained from the model.
    \item Steps 2 and 3 are repeated a number M of times until the required number of models are built.
    \item The final boost prediction is the additive sum of all previous made by the models.
\end{enumerate}

The hyperparameters used to build our GBRegressor were the following:
\begin{itemize}
    \item[--] the loss function used is the \textit{squared error};
    \item[--] the learning rate, which defines the contribution of each tree, is set to 0.01;
    \item[--] the number of boosting stages is set to 10000;
    \item[--] the function used to measure the quality of a split is the \textit{friedman\_mse}, or the mean squared error with the improvement by Friedman;
    \item[--] the minimum number of samples required to split an internal node is set to 2;
    \item[--] the maximum depth of the trees is set to \textit{None}, so the nodes are expanded until all leaves are pure or until all leaves contain less than \textit{min\_samples\_split} samples;
    \item[--] the number of feature to consider when looking for the best split hyperparameter is set to \textit{None}, so that the max features to consider is equal to the number of features available.
\end{itemize}

\section{Sampling techniques}
\label{sec:MCMC}

This section introduces the techniques used in our work to perform cosmological parameters inference using the Pantheon+SH0ES type Ia Supernovae dataset. As we said, the methods used in our project are Monte Carlo Markov chain (MCMC) and Nested Sampling.
MCMC is a probabilistic method that explores the parameter space by generating a sequence of samples, where each sample is a set of parameter values. The core of the method is related to the Markov property, which means that the next state in the sequence depends only on the current state \citep{norris1998markov}.
In the context of cosmological parameter inference, MCMC is often used to sample the posterior distribution of parameters given observational data. It explores the parameter space by creating a chain of samples, with the density of samples reflecting the posterior distribution. Through analysis of this chain, one can estimate the most probable values and uncertainties for cosmological parameters. 
In our work, we used two versions of Nested Sampling: the standard and the dynamic version, which is a slight variation of the former. Nested Sampling is a technique used for Bayesian evidence calculation and parameter estimation and involves enclosing a shrinking region of high likelihood within the prior space and iteratively sampling points from this region. Nested Sampling was developed to estimate the marginal likelihood, but it can also be used to generate posterior samples, and it can potentially work on harder problems where standard MCMC methods may get stuck.
Dynamic Nested Sampling is an extension of Nested Sampling that adapts the sampling strategy during the process. It starts with a high likelihood region and dynamically adjusts the sampling to focus on regions of interest. This method is advantageous for exploring complex and multimodal parameter spaces, which can occur in cosmological models. 
At the end of each technique, we evaluated its performance by calculating the Bayesian Information Criterion \citep{schwarz1978estimating} (BIC) and the Akaike Information Criterion \citep{akaike1974new} (AIC).

\subsection{Monte Carlo Markov chain (MCMC)}

Bayesian inference treats probability as a measure of belief, with parameters treated as random variables influenced by data and prior knowledge. The goal is to combine prior information with observational data to refine our estimates and obtain a posterior distribution that summarises everything we know about the parameters \citep{bolstad2009understanding}.

The posterior distribution is given by:
\begin{equation}
    P(\Theta|D) \propto P(D|\Theta) \cdot \pi(\Theta),
\end{equation}

where $P(D|\Theta)$ is the likelihood, and $\pi(\Theta)$ is the prior. Since this distribution is often difficult to calculate analytically, MCMC provides a way to approximate it by generating a chain of parameter sets that converge to the posterior distribution over time. This allows us to estimate key cosmological parameters, even in high-dimensional spaces where standard methods struggle.

\begin{table*}
\centering
\small
\begin{tabular}{lcccccccccc}
\toprule
\multirow{2}{*}{Model} & \multicolumn{2}{c}{$H_0$} & \multicolumn{2}{c}{$\Omega_m$} & \multicolumn{2}{c}{$w$} \\
& Mean & Std & Mean & Std & Mean & Std \\
\midrule
$\Lambda$CDM & 70.0 & 0.1 & 0.3 & 0.05 & -1.0 & 0.1 \\
\midrule
& \multicolumn{2}{c}{$H_0$} & \multicolumn{2}{c}{$\Omega_m$} & \multicolumn{2}{c}{$w_0$} & \multicolumn{2}{c}{$w_z$} \\
& Mean & Std & Mean & Std & Mean & Std & Mean & Std \\
\midrule
Linear Redshift & 70.0 & 0.1 & 0.3 & 0.05 & -1.0 & 0.1 & -0.1 & 0.1 \\
\midrule
& \multicolumn{2}{c}{$H_0$} & \multicolumn{2}{c}{$\Omega_m$} & \multicolumn{2}{c}{$w_0$} & \multicolumn{2}{c}{$w_a$} \\
& Mean & Std & Mean & Std & Mean & Std & Mean & Std \\
\midrule
CPL & 70.0 & 0.1 & 0.3 & 0.05 & -1.0 & 0.1 & -0.5 & 0.1 \\
\midrule
& \multicolumn{2}{c}{$H_0$} & \multicolumn{2}{c}{$\Omega_m$} & \multicolumn{2}{c}{$w_0$} & \multicolumn{2}{c}{$w_a$} \\
& Mean & Std & Mean & Std & Mean & Std & Mean & Std \\
\midrule
Squared Redshift & 70.0 & 0.1 & 0.3 & 0.05 & -1.0 & 0.1 & -0.1 & 0.1 \\
\midrule
& \multicolumn{2}{c}{$H_0$} & \multicolumn{2}{c}{$\Omega_m$} & \multicolumn{2}{c}{$b$} & \multicolumn{2}{c}{$\alpha$} \\
& Mean & Std & Mean & Std & Mean & Std & Mean & Std \\
\midrule
Generalized CG & 70.0 & 0.1 & 0.3 & 0.05 & -1.0 & 0.1 & 0.2 & 0.1 \\
\midrule
& \multicolumn{2}{c}{$H_0$} & \multicolumn{2}{c}{$\Omega_m$} & \multicolumn{2}{c}{$b$} & \multicolumn{2}{c}{$b_s$} & \multicolumn{2}{c}{$\alpha$} \\
& Mean & Std & Mean & Std & Mean & Std & Mean & Std & Mean & Std \\
\midrule
Modified CG & 70.0 & 0.1 & 0.3 & 0.05 & -1.0 & 0.1 & 0 & 0.1 & 0.2 & 0.1 \\
\bottomrule
\end{tabular}
\caption{Initial conditions for MCMC.}
\label{tab:MCMC}
\end{table*}

\subsubsection{Markov chains}

MCMC is based on the concept of a Markov chain, where each state depends only on the previous one (the so-called Markov property):
\begin{equation}
P(x^{(i)} | x^{(i-1)}, ..., x^{(1)}) = P(x^{(i)} | x^{(i-1)}).
\end{equation}
The process is designed to ensure that the chain converges to the target posterior distribution after a number of steps. A key requirement is the detailed balance condition:
\begin{equation}
p(x^{(i)})T(x^{(i-1)} | x^{(i)}) = p(x^{(i-1)})T(x^{(i)} | x^{(i-1)}).
\end{equation}
This ensures that the chain remains in the desired distribution, allowing accurate parameter estimates.

\subsubsection{The Metropolis-Hastings (M-H) algorithm}

The simplest and most commonly used MCMC algorithm is the M-H method \citep{metropolis1953equation, mackay2003information, gregory2005bayesian, hogg2010data}. The iterative procedure is the following:
\begin{enumerate}
    \item given a position $X(t)$ sample a proposal position $Y$ from the transition distribution $Q(Y;X(t))$;
    \item accept this proposal with probability
        \begin{equation}
            min\left(1, \frac{p(X|D)}{p(X(t)|D)}\frac{Q(X(t);Y)}{Q(Y;X(t))}\right),
        \end{equation}
\end{enumerate}
where $D$ is a set of observations.
If accepted, the new position will be $X(t+1) = Y$; otherwise, it remains at $X(t+1) = X(t)$.

This algorithm converges to a stationary distribution over time, but there are alternative methods that can achieve faster convergence depending on the problem.

\subsubsection{The stretch move}

The stretch move algorithm, proposed by \citet{goodman2010insem}, is an affine-invariant method that outperforms the standard M-H algorithm by producing independent samples with shorter autocorrelation times. It works by generating an ensemble of $K$ walkers, where the proposal for a walker is based on the current positions of the others.

The new position for walker $X_k$ is proposed using:

\begin{equation}
    X_k(t) \rightarrow Y = X_j + Z[X_k(t) - X_j],  \label{eq:X_k}
\end{equation}
where $Z$ is a random variable. This method ensures detailed balance and faster convergence, making it suitable for high-dimensional parameter spaces.

\subsubsection{Our implementation of Monte Carlo Markov chain}
In our work, we applied Monte Carlo Markov chain (MCMC) to both the original dataset, which includes measured redshifts and distance modulus, and the predicted dataset, which includes measured redshift and predicted distance modulus obtained from our ensemble learning model. The hyperparameters used in our analysis are as follows:
\begin{itemize}
    \item[--] The number of walkers is set to 100.
    \item[--] The move is the previously discussed StretchMove.
    \item[--] The number of steps varies depending on the original and predicted dataset, with 1000 steps for the former and, to account for the additional uncertainty of the machine learning models, 2500 steps for the latter.
    \item[--] The number of initial steps of each chain discarded as \textit{burn-in} is set to 100.
    \item[--] The initial positions of the walkers are randomly generated around the initial values for the cosmological parameters investigated with specified standard deviations (See Table \ref{tab:MCMC}). 
\end{itemize}

\subsection{Nested Sampling}

Modern astronomy often involves inferring physical models from large data sets. This has shifted the standard statistical framework from frequentist approaches such as Maximum Likelihood Estimation \citep{fisher1922mathematical} (MLE) to Bayesian methods, which estimate the distribution of parameters consistent with the data and prior knowledge. While Monte Carlo Markov chain (MCMC) is widely used for Bayesian inference \citep{brooks2011handbook, sharma2017markov}, it can struggle with complex, multimodal distributions and doesn't directly estimate the model evidence needed for model comparison \citep{plummer2003jags, foreman2013emcee, carpenter2017stan}.

Nested Sampling \citep{skilling2006nested} provides a solution by focusing on both posterior sampling and evidence estimation. It samples from nested regions of increasing likelihood, allowing more effective exploration of complex parameter spaces. The final set of samples, combined with their importance weights, helps to generate posterior estimates while also providing a way to compute evidence for model comparison.

Unlike MCMC, which directly estimates the posterior $P(\Theta)$, Nested Sampling decomposes the problem by:

\begin{enumerate}
    \item Splitting the posterior into simpler distributions.
    \item Sampling sequentially from each distribution.
    \item Combining the results to estimate the overall posterior and evidence.
\end{enumerate}

The goal is to compute the evidence $\mathcal{Z}$, given by:

\begin{equation}
\mathcal{Z} = \int _{\Omega_\Theta} \mathcal{L}(\Theta)\pi(\Theta)d\Theta = \int _0 ^1 \mathcal{L}(X)dX, \label{eq:Z2}
\end{equation}

where $\mathcal{L}(X)$ defines iso-likelihood contours outlining the prior volume $X$.

This procedure allows Nested Sampling to handle complex, high-dimensional problems that may be difficult for MCMC.

\begin{figure}
    \centering
    \includegraphics[width=\columnwidth]{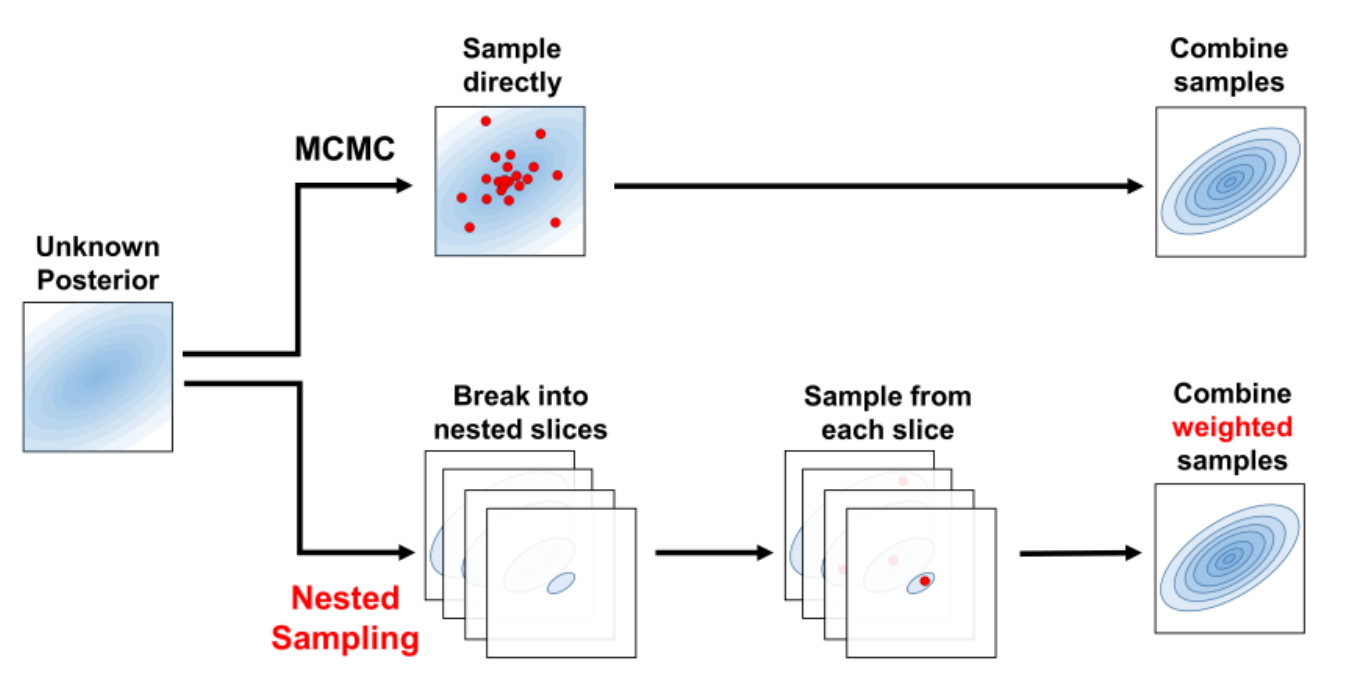}
    \caption[MCMC vs Nested Sampling]{\small While MCMC methods attempt to generate samples directly from the posterior, Nested Sampling instead breaks up the posterior into many nested “slices”, generates samples from each of them, and then recombines the samples to reconstruct the original distribution using the appropriate weights \citep{speagle2020dynesty}.}
    \label{fig:MCMC vs Nested Sampling}
\end{figure}

Figure \ref{fig:MCMC vs Nested Sampling} llustrates the difference between MCMC methods and nested sampling, where MCMC generates samples directly from the posterior, whereas nested sampling breaks the posterior into nested slices, samples from each, and then combines them to reconstruct the original distribution with appropriate weights.

\subsubsection{Stopping criterion} \label{sec:Stopping criterion NS}

Nested Sampling typically stops when the estimated remaining evidence is below a certain threshold \citep{keeton2011statistical, higson2018sampling}. The stopping condition is:

\begin{equation}
    \Delta \ln \hat{\mathcal{Z}}_i < \epsilon, \label{eq:stoppingcriterion}
\end{equation}

where $\epsilon$ is a user-defined tolerance that indicates how much evidence remains to be integrated. In the Python library \texttt{dynesty} \citep{speagle2020dynesty, higson2019dynamic}, this tolerance is typically set to 1\%.

\subsubsection{Evidence and Posterior}

Once the sampling is complete, the evidence $\mathcal{Z}$ is estimated by numerical integration, typically using the trapezoidal rule:

\begin{equation}
    \hat{\mathcal{Z}} = \sum_{i=1}^{N+K} \frac{1}{2} [\mathcal{L}(\Theta_{i-1}) + \mathcal{L}(\Theta_i)] \times (\hat{X}_{i-1} - \hat{X}_i).
\end{equation}

The posterior $P(\Theta)$ is then calculated by weighting the samples based on their probabilities and the volume they represent.

\subsubsection{Benefits and drawbacks of Nested Sampling} \label{sec:Drawbacks of NS}

Nested Sampling has several advantages over traditional MCMC methods:

\begin{enumerate}
	\item It can estimate both the evidence $\mathcal{Z}$ and the posterior $P(\Theta)$, whereas MCMC generally focuses only on the posterior \citep{lartillot2006computing, heavens2017marginal}.
	\item It performs well with multimodal distributions, which can be difficult for MCMC to handle.
	\item The stopping criteria in Nested Sampling are based on evidence estimation, providing a more natural endpoint than MCMC, which uses sample size-based criteria.
	\item Nested Sampling starts integrating the posterior from the prior, allowing it to explore the parameter space smoothly without having to wait for convergence, unlike MCMC \citep{gelman1992inference, vehtari2021rank}. 
\end{enumerate}

However, Nested Sampling has limitations:

\begin{enumerate}
  \item It often samples from uniform distributions, which can limit flexibility when dealing with more complex priors.
  \item Runtime increases with the size of the prior, making it less efficient if the prior is large but doesn't significantly affect the posterior.
  \item The integration rate remains constant throughout the process, so it doesn't allow priorisation of the posterior over the evidence, even as the number of live points increases.
\end{enumerate}

\subsubsection{Bounding Distributions}

In Nested Sampling, the current live points are used to estimate the shape and size of regions in parameter space, which helps to guide sampling more efficiently.

We used the multiple ellipsoids method, where:

\begin{enumerate}
    \item A bounding ellipsoid is first constructed to enclose all live points.
    \item K-means\footnote{K-means is a clustering algorithm that categorizes data into K clusters by iteratively assigning data points to the cluster with the closest centroid, aiming to minimise the variance within each cluster. This process continues until convergence, resulting in K clusters with centroids that represent the centre of each cluster's data points.} clustering is used to divide the points into clusters, with new ellipsoids constructed around each cluster.
    \item This process continues iteratively until no further decomposition is required.
\end{enumerate}

\subsubsection{Sampling Methods}

After constructing a bounds distribution, \texttt{dynesty} proceeds to generate samples conditioned on those bounds. We utilized a uniform sampling method in our work.

The general procedure for generating uniform samples from overlapping bounds is as follows \citep{feroz2008multimodal}:

\begin{enumerate}
    \item Select a boundary with probability proportional to its volume.
    \item Sampling a point uniformly from the selected boundary.
    \item Accept the point with a probability inversely proportional to the number of overlapping boundaries.
\end{enumerate}

This approach ensures that samples are drawn efficiently from the defined bounds, maximising the likelihood of finding points in high probability regions of the parameter space.

\subsubsection{Dynamic Nested Sampling}

In Sect. \ref{sec:Drawbacks of NS}, we identified three main limitations of standard Nested Sampling:

\begin{enumerate}
    \item The need for a prior transform.
    \item Increased running time for larger priors.
    \item A constant rate of posterior integration.
\end{enumerate}

While the first two are inherent to the Nested Sampling method, the third limitation can be addressed by adjusting the number of live points during the run. This approach, known as Dynamic Nested Sampling \citep{higson2019dynamic}, allows the algorithm to focus more on either the posterior $(P(\Theta))$ or the evidence ($\mathcal{Z}$), providing flexibility that standard Nested Sampling lacks.

The key idea is to increase the number of live points $K$ in areas where more detail is needed and reduce it where faster exploration is sufficient. This makes it more efficient to deal with complex parameter spaces without sacrificing accuracy in posterior or evidence estimation.

The number of live points $K(X)$ as a function of the prior volume $X$ is guided by an importance function $I(X)$, which determines how resources are allocated during sampling. In \texttt{dynesty}, this importance function is defined as:

\begin{equation}
    I(X) = f^P I^P(X) + (1 - f^P) I^{\mathcal{Z}}(X), 
\end{equation}
where $f^P$ is the relative importance assigned to estimating the posterior.

Posterior Importance $I^P(X)$ is proportional to the probability density of the importance weight $p(X)$, meaning more live points are allocated in regions with high posterior mass. On the other hand, evidence importance $I^{\mathcal{Z}}$ focuses on regions where there is uncertainty in integrating the posterior, ensuring accurate evidence estimation.

By varying the number of live points based on these importance functions, Dynamic Nested Sampling strikes a balance between efficient sampling and accurate evidence estimation, improving performance in complex scenarios.

\subsubsection{Our implementation of Static and Dynamic Nested sampling}

As with the MCMC method, we applied Static and Dynamic Nested Sampling to both the original and predicted datasets. The hyperparameters used are the same for both the sampling methods and are the following:
\begin{itemize}
    \item[--] The number of live points varies depending on the original and predicted dataset, with 1000 steps for the former and, to account for the additional uncertainty of the machine learning models, 2500 steps for the latter.
    \item[--] The bounding distribution used is the multi ellipsoids.
    \item[--] The sampling method used is  uniform.
    \item[--] The maximum number of iterations, as the number of likelihood evaluations, is set to \textit{no limit}. Iterations will stop when the termination condition is reached.
    \item[--] The \textit{dlogz} value, which sets the $\epsilon$ of the termination condition (Eq. \ref{eq:stoppingcriterion}), is set to 0.01.
\end{itemize}

\subsection{Information Criteria}

Let us consider now two statistical criteria in order to compare our MCMC and Nested Sampling results:
\begin{itemize}
    \item[--] The Akaike Information Criterion (AIC), defined as
        \begin{equation}
            AIC = -2\ln\mathcal{L} + 2k,
        \end{equation}
        where $\mathcal{L}$ is the maximum likelihood and $k$ is the number of parameters in the model. The optimal model is the one that minimises the AIC, since it provides an estimate of a constant plus the relative difference between the unknown true likelihood function of the trained sampler and the fitted likelihood function of the cosmological model. Therefore, a lower AIC indicates that the model is closer to the true underlying likelihood.
    \item[--] The Bayesian Information Criterion (BIC) defined as:
        \begin{equation}
            BIC = -2\ln\mathcal{L} + k\ln N,
        \end{equation}
        where \(N\) is the number of data points used in the fit.
        The BIC serves as an estimate of a function related to the posterior probability of a model being the true model within a Bayesian framework. Therefore, a lower BIC indicates that a model is deemed more likely to be the true model.
\end{itemize}

\section{Results}
\label{sec:RES}

As mentioned before, our work can be divided into two main sections: the first one, where we operate on the original SNe type Ia dataset; the second one, where we use the predicted distance modulus dataset after feature selection methods and an ensemble model. 
In particular, as discussed earlier, we used three feature selection methods to build the predicted dataset:
\begin{itemize}
    \item[--] Random Forest: here we have taken the 18 most important features out of the 70 total. The features considered are: \textit{m\_b\_corr}, \textit{mB}, \textit{zCMB}, \textit{x0}, \textit{zHEL}, \textit{zHD}, \textit{std\_flux}, \textit{m\_b\_corr\_err\_VPEC}, \textit{MU\_SH0ES\_ERR\_DIAG}, \textit{m\_b\_corr\_err\_DIAG}, \textit{skew}, \textit{NDOF}, \textit{percent\_amplitude}, \textit{DEC}, \textit{biasCors\_m\_b\_COVSCALE}, \textit{fpr35}, \textit{K}, \textit{COV\_c\_x0}. 
    \item[--] Boruta: here we have taken the \textit{confirmed} features. The 13 accepted features employed are: \textit{m\_b\_corr}, \textit{mB}, \textit{x0}, \textit{zCMB}, \textit{zHEL}, \textit{zHD}, \textit{std\_flux}, \textit{m\_b\_corr\_err\_VPEC}, \textit{MU\_SH0ES\_ERR\_DIAG}, \textit{m\_b\_corr\_err\_DIAG}, \textit{percent\_amplitude}, \textit{DEC}, \textit{MWEBV}.
    \item[--] SHAP: as with the Random Forest, also here we have taken the 18 most important features. The features selected are: \textit{m\_b\_corr}, \textit{mB}, \textit{zCMB}, \textit{x0}, \textit{zHEL}, \textit{zHD}, \textit{std\_flux}, \textit{m\_b\_corr\_err\_VPEC}, \textit{MU\_SH0ES\_ERR\_DIAG}, \textit{m\_b\_corr\_err\_DIAG}, \textit{percent\_amplitude}, \textit{DEC}, \textit{NDOF}, \textit{biasCors\_m\_b\_COVSCALE}, \textit{fpr35}, \textit{skew}, \textit{RA}, \textit{COV\_c\_x0}.
\end{itemize}

To summarise and highlight the differences between the three parameter spaces used, Fig. \ref{fig:Feature Importance Comparison} below shows the Random Forest and SHAP importance, on a logarithmic scale, of the features selected by at least one of the techniques, together with the Boruta classification. 

\begin{figure}
	\centering
	\includegraphics[width=\columnwidth]{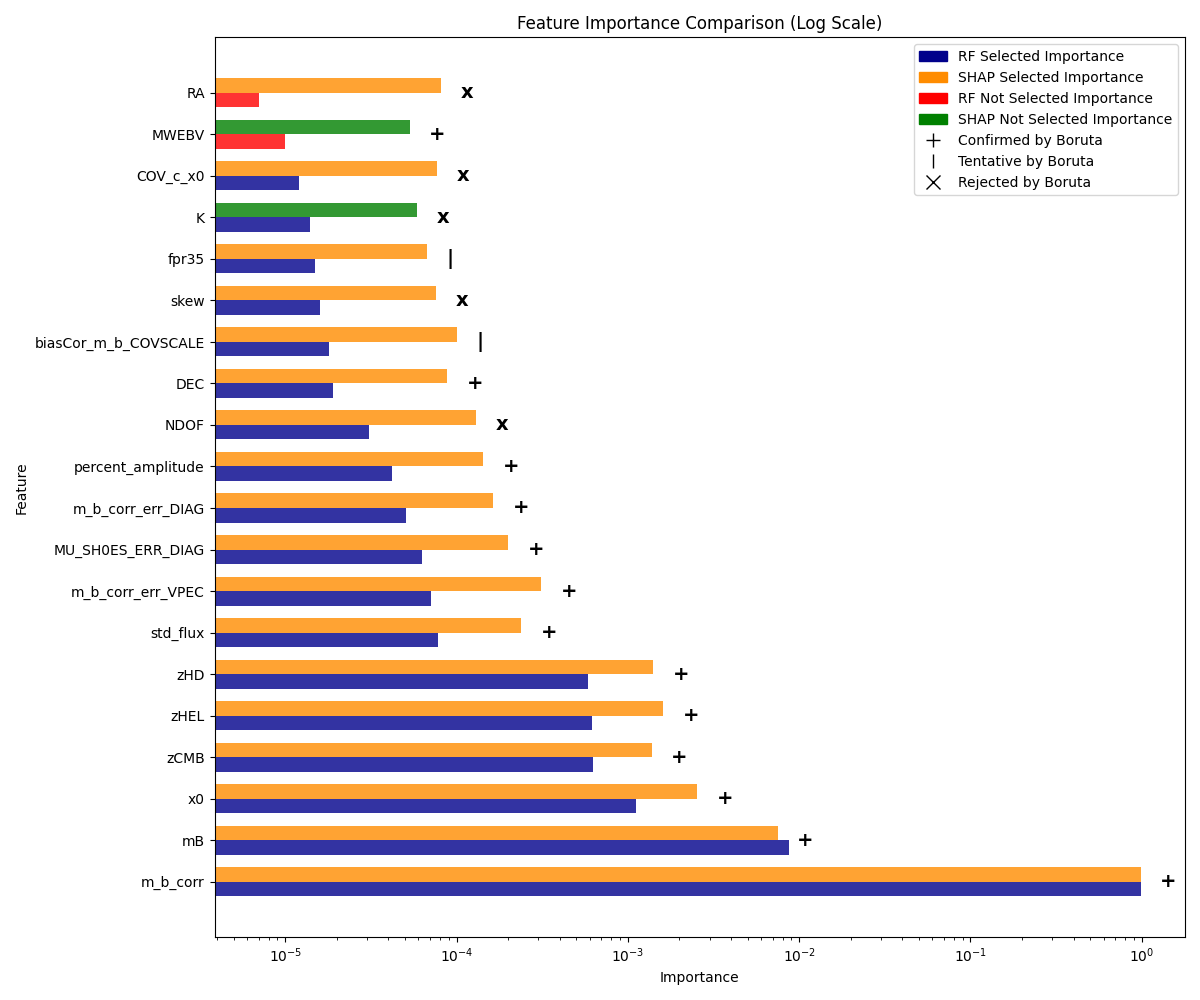}
	\caption[Feature Importance Comparison]{\small Feature Importance Comparison: The graph uses dark blue bars to show the importance of features selected by the RF model, while dark orange bars show the importance of features selected by SHAP. Red bars represent the importance of features not selected by the RF model, and green bars represent the importance of features not selected by SHAP. The symbols indicate the Boruta classification: $\mathbf{+}$ for confirmed, $\mathbf{|}$ for tentative, and $\mathbf{x}$ for rejected features. More information on all features (Pantheon+SH0ES dataset and additions) can be found in the Appendix.}
	\label{fig:Feature Importance Comparison}
\end{figure}

We also performed a 'base' case study where all 70 features were used to predict distance moduli with an ensemble model. After each MCMC and Nested Sampling iteration, BIC and AIC have been  computed. The final cosmological parameters and their uncertainties have been obtained as the mean of the three methods used.
This work is developed for all the previously introduced six cosmological models, thanks to the \texttt{Astropy} Python package \citep{price2022astropy}. A summary of the results is provided in the Appendix. This includes corner plots obtained by each technique and a table showing the key results, such as BIC and AIC scores, for each method. 

In the next plots we will indicate with 'OR' the case of the original dataset, with 'ALL' the case where no feature selection is done (i.e. all features are used), with 'RF' the case where Random Forest is used as feature selection technique, with 'BOR' the case where Boruta technique is used and with 'SHAP' the case where SHAP is used as feature selection method. In addition,'MCMC', 'SNS' (Static Nested Sampling) and 'DNS' (Dynamic Nested Sampling) indicate the different sampling techniques.

\begin{figure}
	\centering
	\includegraphics[width=\columnwidth]{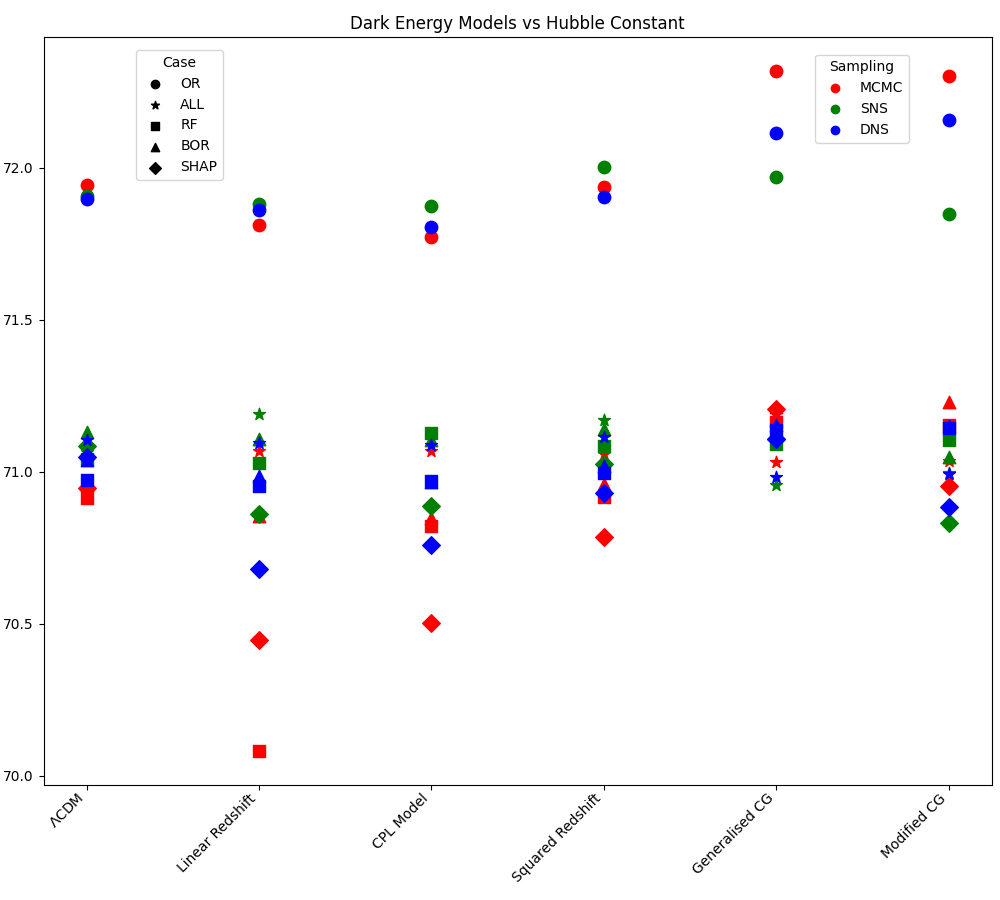}
	\caption[Hubble Constant Values]{Hubble Constant Values.}
	\label{fig:Hubble Constant Values}
\end{figure}

\begin{figure}
	\centering
	\includegraphics[width=\columnwidth]{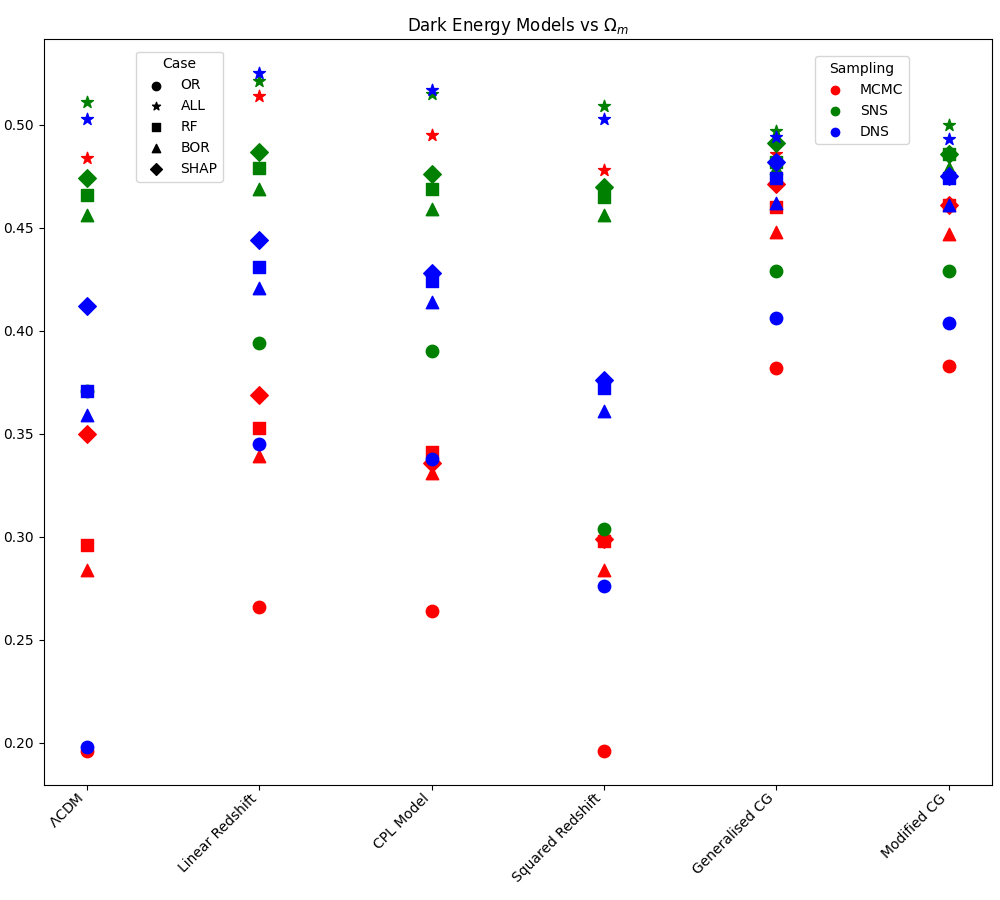}
	\caption[Matter Density Values]{Matter Density Values.}
	\label{fig:Matter Density Values}
\end{figure}

\begin{figure}
	\centering
	\includegraphics[width=\columnwidth]{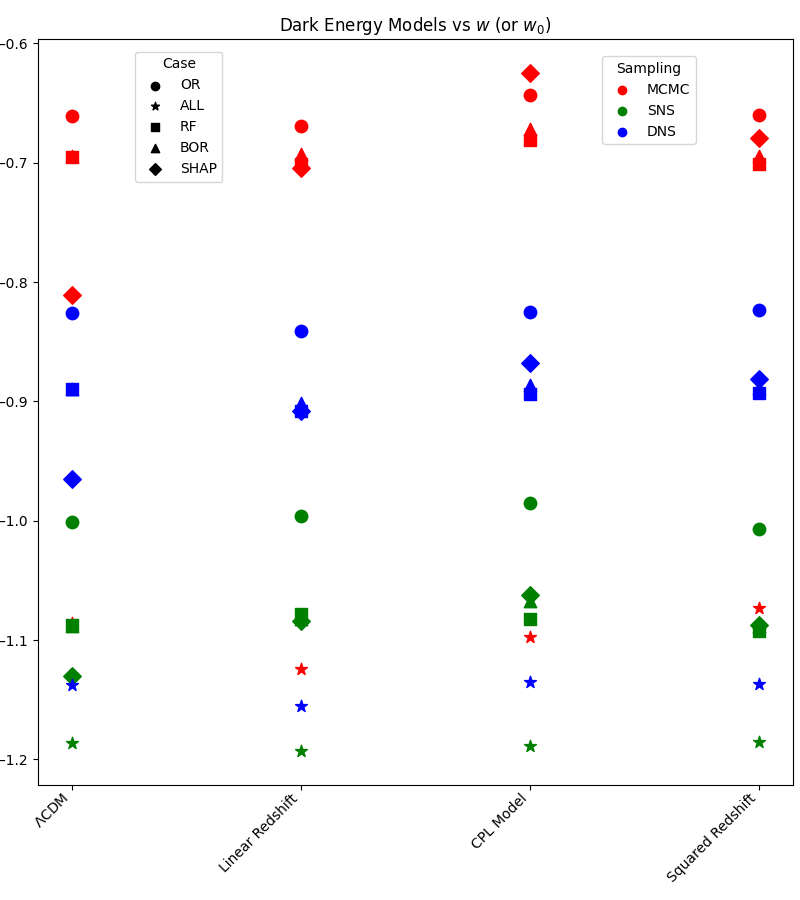}
	\caption[$w$ ($w_0$) Values]{$w$ ($w_0$) Values.}
	\label{fig:$w$ ($w_0$) Values}
\end{figure}

\begin{figure}
	\centering
	\includegraphics[width=1\columnwidth]{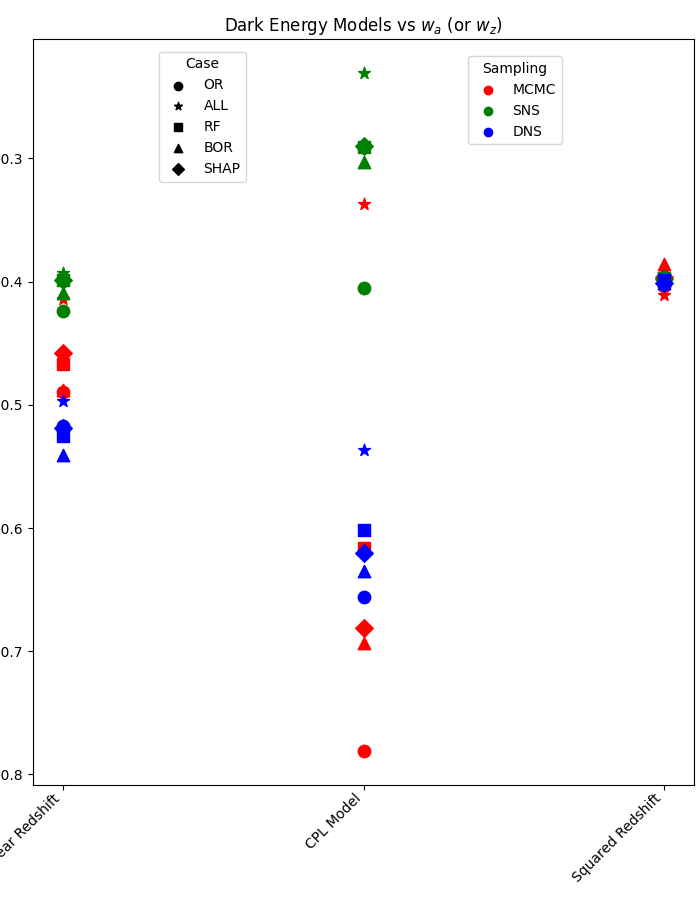}
	\caption[$w_a$ ($w_z$) Values]{$w_a$ ($w_z$) Values.}
	\label{fig:$w_a$ ($w_z$) Values}
\end{figure}

\begin{figure*}
    \centering
    \begin{minipage}[b]{0.25\textwidth}
        \centering
        \includegraphics[width=\textwidth]{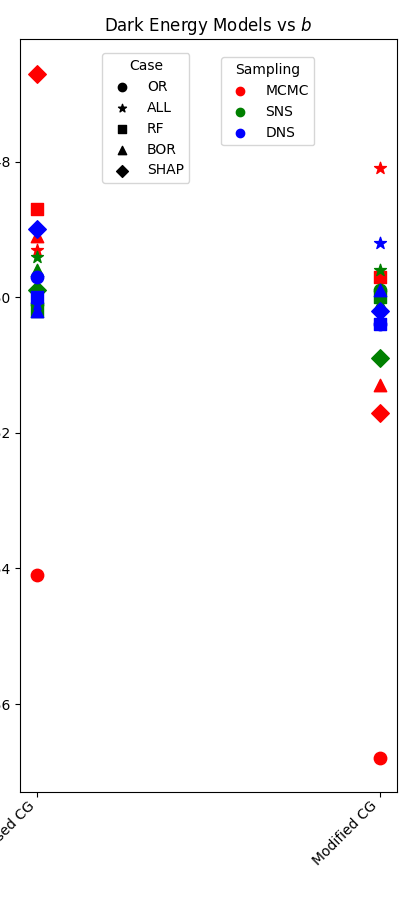}
        \caption{$b$ Values}
        \label{fig:b_Values}
    \end{minipage}
    \hfill
    \begin{minipage}[b]{0.25\textwidth}
        \centering
        \includegraphics[width=\textwidth]{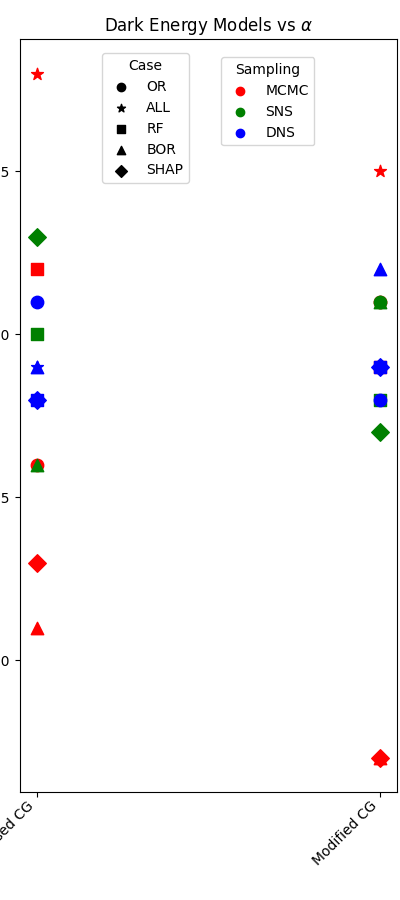}
        \caption{$\alpha$ Values}
        \label{fig:alpha_Values}
    \end{minipage}
    \hfill
    \begin{minipage}[b]{0.25\textwidth}
        \centering
        \includegraphics[width=\textwidth]{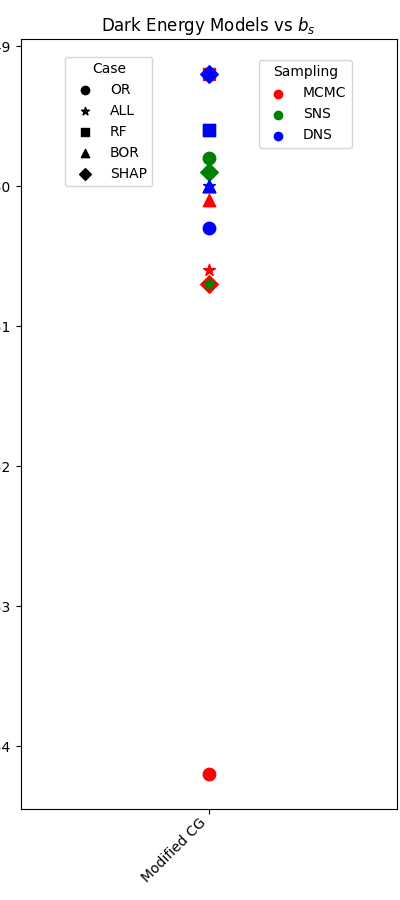}
        \caption{$b_s$ Values}
        \label{fig:bs_Values}
    \end{minipage}
\end{figure*}

\begin{figure*}
    \begin{minipage}[b]{0.49\textwidth}
        \centering
        \includegraphics[width=\linewidth]{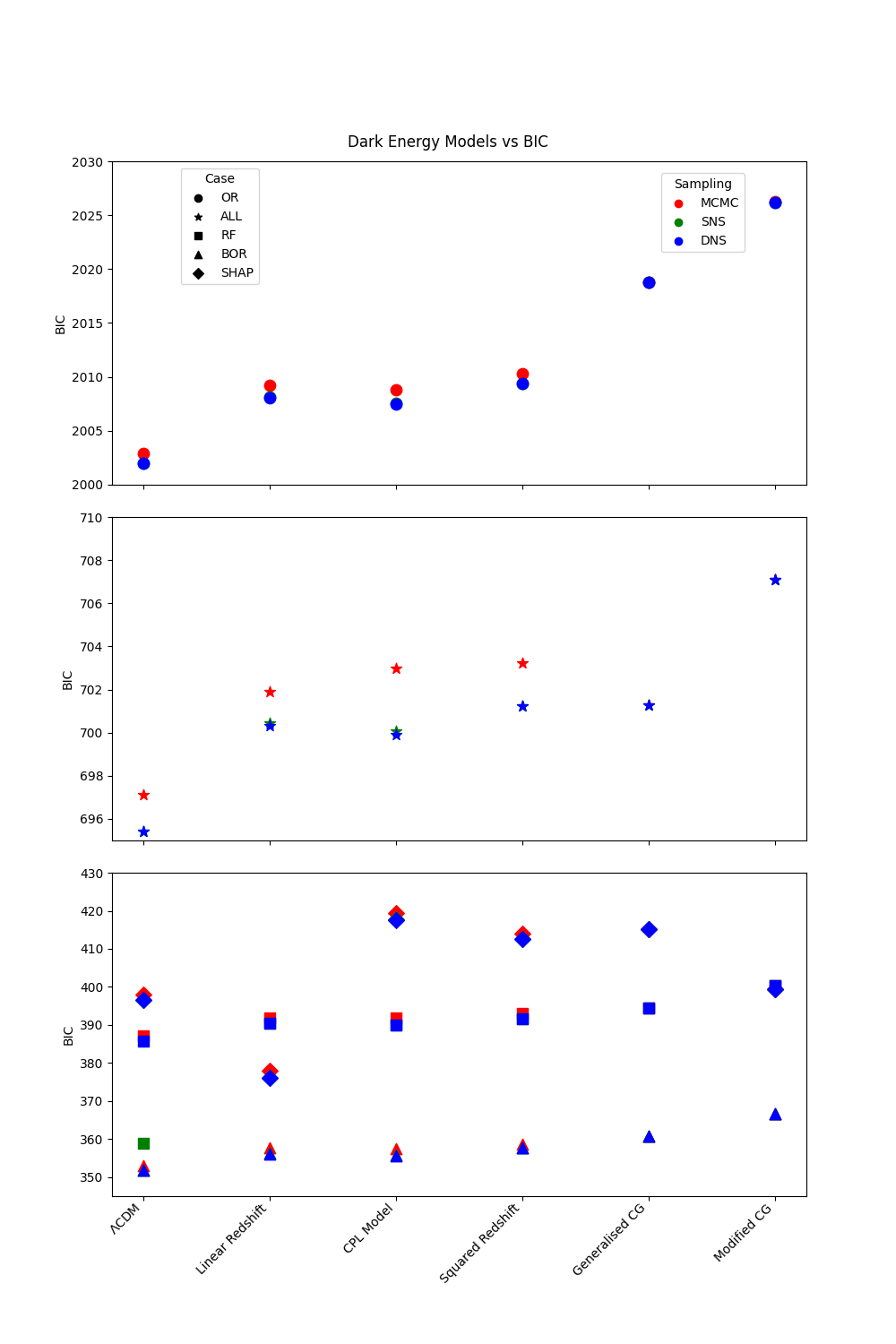}
        \caption[BIC Values]{BIC Values.}
        \label{fig:BIC_Values}
    \end{minipage}
    \begin{minipage}[b]{0.49\textwidth}
        \centering
        \includegraphics[width=\linewidth]{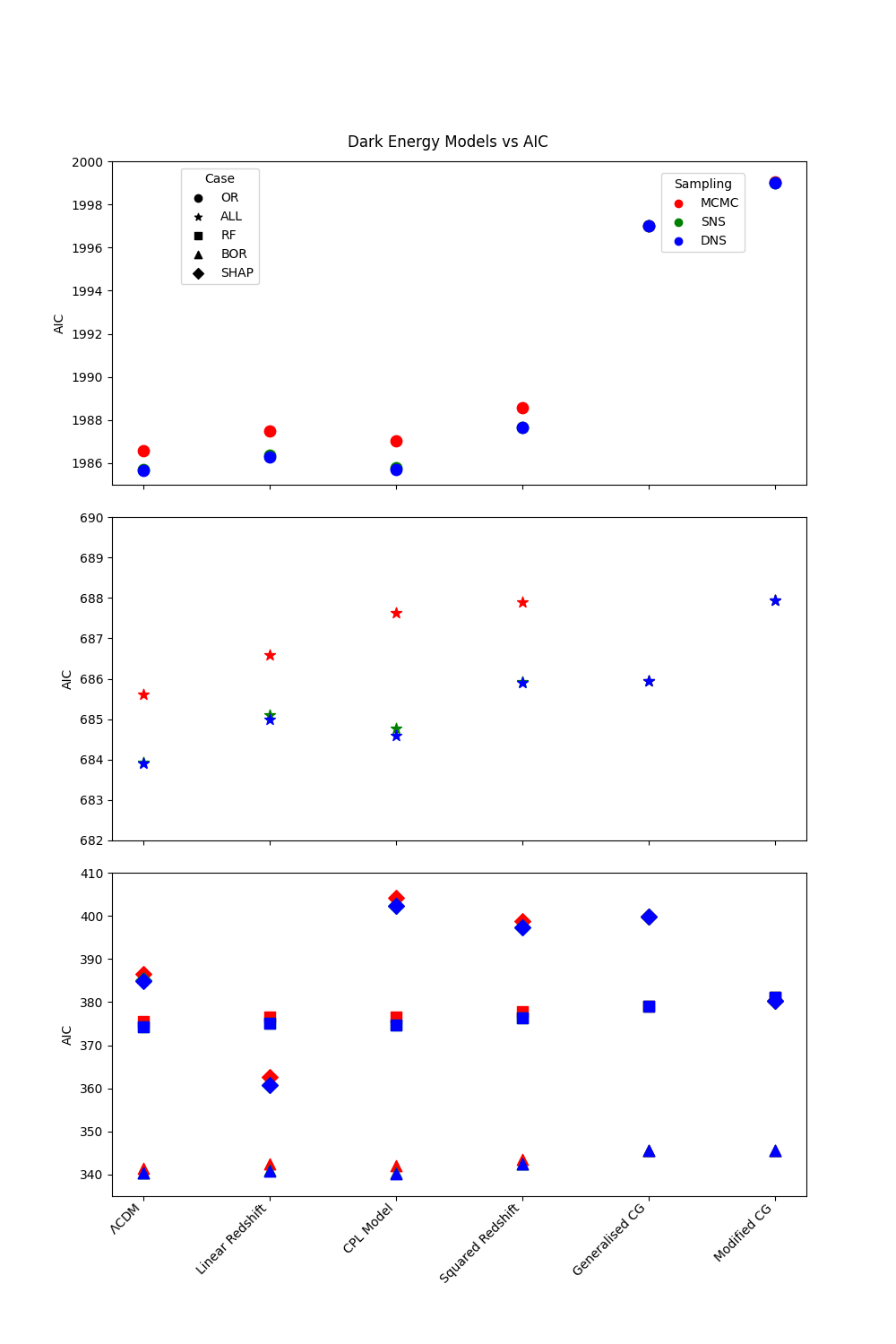}
        \caption[AIC Values]{AIC Values.}
        \label{fig:AIC_Values}
    \end{minipage}
\end{figure*}

Figure \ref{fig:Hubble Constant Values} shows the derived values of the Hubble constant H$_0$ for each cosmological model under different feature selection methods. The plot contrasts the results obtained from the original dataset with those obtained using feature selection techniques along with different sampling techniques such as MCMC, Static Nested Sampling (SNS) and Dynamic Nested Sampling (DNS). In particular, the results from the original dataset tend to be higher compared to all other values found in the second part. The opposite trend is seen in Fig. \ref{fig:Matter Density Values}, which compares the matter density parameter. The results show significant deviations when no feature selection is applied, emphasising the importance of selecting relevant features to avoid skewed or biased estimates. In addition, the Generalised and Modified Chaplygin Gas models show a different behaviour compared to other models, with generally higher values, further emphasising their divergence from standard cosmological models.
In the Fig. \ref{fig:$w$ ($w_0$) Values}, where the evolution of the equation of state parameter $w_0$ is shown for the different models, the values seem to be related to the particular sampling technique used, with the exception of the case where no feature selection was applied, which is an alarming sign for its performance. Furthermore, in the Fig. \ref{fig:$w_a$ ($w_z$) Values} presents the analysis of the equation of state parameter $w_a$ (or $w_z$). A notable finding is that the Linear and Squared Redshift models give more or less the same results, while the CPL model covers a much wider range of values. In addition, Figs. \ref{fig:b_Values}--\ref{fig:bs_Values} show the parameters specific to the Chaplygin gas models, $b$, $\alpha$ and $b_s$ respectively.

From the Figs. \ref{fig:BIC_Values} and \ref{fig:AIC_Values}, which show the values of BIC and AIC respectively for the analysed models, we can draw some interesting conclusions. Firstly, the values of BIC and AIC are much higher in the original dataset with respect to the other four cases, but this is due to the difference in the size of the dataset used, complete for the first scenario and 20\% for the others, which leads to much larger penalty terms in the final values of the Information Criteria. Secondly, among the scenarios of the second part of the study, the case where no feature selection has been applied, has the highest values in BIC and AIC, which is representative of the fact that this is the worst case analysed, because not only  we do not obtain a better performance, but we also have the higher complexity in the model. Among the three cases analysed, Boruta clearly has the lowest Information Criteria values and therefore the better sampling performance. Looking at the models, it is interesting to note that from the original dataset scenario to those where the feature selection has been applied, the Chaplygin Gas models have the greatest increase in performance compared to the other models. Finally, remaining in the three cases of feature selection, the model that has the lowest mean values of BIC and AIC is the Linear Redshift one, but this may be due to the low redshift of our dataset, which favours this model.

\section{Discussion and Conclusions}
\label{sec:CONCL}

In this work we have performed a test of six dark energy models with recent observations of Type Ia Supernovae. First, we tested each model by inferring its cosmological parameters by using Monte Carlo Markov chain, Static Nested Sampling and Dynamic Nested Sampling. Secondly, we tried a different approach using machine learning. We built a regression model where the distance modulus of each supernova, the crucial data for inferring the cosmological parameters, was computed by the machine learning model, thanks to the other available features. In fact, we have not only relied on the features provided by the original dataset \citep{scolnic2022pantheon+}, but we have extended it by several features, bringing the total number to 74. The machine learning model used to compute the distance moduli is an ensemble model composed by four models: the MultiLayer Perceptron, the k-Nearest Neighbours, the Random Forest Regressor and the Gradient Boosting model. In order to improve the performance of our ensemble learning model, we applied different feature selection techniques, emphasising the importance of a data-driven approach. We have inferred the cosmological parameters of each model in four different cases: a case where no feature selection is applied (a sort of 'base' case), a case where the first 18 features selected by the Random Forest are used to infer the distance moduli, a case where the feature selection method used is Boruta, and finally the case where the features used are the first 18 selected by SHAP. For every case, we repeated the process done in the first part of the study, or the use of MCMC and Nested Sampling, to infer the cosmological parameters for each model. By incorporating feature selection methods, we ensured that our models focused on the most relevant and informative features, thereby improving the robustness of the distance modulus predictions.

In the first phase of our study, the $\Lambda$CDM parameters were found to be consistent with established observations, confirming its status as a robust standard cosmological model. While the introduction of new parameters in the linear, squared redshift and CPL models led to slight deviations, the overall parameter values remained relatively similar across the different parameterisations. Instead, the Generalised and Modified Chaplygin Gas models showed significant deviations, especially in the matter density parameter, making them the worst performing of the six models.

Moving to the second part of our work, it is important to  point out the results of the feature selection processes. The analysis shows that the most important feature by a significant margin is \textit{m\_b\_corr}, which represents the Tripp1998 corrected/standardised $m_b$ magnitude. Following at some distance are \textit{mB} (SALT2 uncorrected brightness, \citet{guy2007salt2}) and \textit{x0} (SALT2 light curve amplitude). Next, in importance, are \textit{zCMB}, \textit{zHEL} and \textit{zHD}, corresponding to the CMB corrected redshift, the heliocentric redshift and the Hubble diagram redshift respectively. While the remaining features are of lesser importance, their contributions are roughly comparable.
It is worth noticing that the selected features are predominantly from the original dataset, with only a few additions made by us, such as \textit{std\_flux}, \textit{percent\_amplitude}, \textit{skew}, \textit{fpr35}, and \textit{K}. This highlights the robustness of the original dataset features in influencing the predictive power of our models. However, the inclusion of additional features by us has provided valuable insights and contributed to the overall effectiveness of the feature selection process.

In the second section of our work, the first result we noticed is the clear difference in the performance of the ensemble model between the case where no feature selection is applied and the three cases where it is present. In the former, the parameters differ significantly from the values found in the first part of the work, but also the values of the information criteria, BIC and AIC, are almost two times the values of the cases where feature selection is applied.
The performances observed for the three feature selection models are quite close, following a similar trend to the first part of the study. In particular, by looking at the values of the most important cosmological parameters, the Generalised and Modified Chaplygin Gas models appear to be slightly less effective than the other four models. Among Random Forest, Boruta and SHAP, the former seems to perform slightly worse, while the other two show comparable results. Furthermore, our analysis reveals an interesting aspect in the estimation of the $w_a$ (or $w_z$) parameter across the dark energy models. The Linear and Squared Redshift parameterisation models give similar estimates, while the CPL model shows a larger variation. In general, the trend across all six models indicates a slightly lower $H_0$ and a slightly higher $\Omega_m$ compared to the values obtained in the first part of the study.
It is worth noting that Boruta stands out as the model with relatively lower information criteria values. It is interesting to note that when looking at the BIC and AIC values, the models that seemed to be by far the worst in the first part of the study, i.e. the Generalised and Modified Chaplygin Gas models, in the case where no feature selection is applied, the result is only confirmed for the Modified Chaplygin Gas, while the Generalised one is among the best models. Instead, for the three cases of feature selection, the opposite happens, with the Generalised model behaving similarly to the CPL and Squared Redshift models, while the Modified model performs better than all these three.
In summary, the feature selection models, especially Boruta, show consistent performance with variations in $H_0$ and $\Omega_m$. The unexpected ranking of the information criteria between the models, which challenges conventional expectations based on theoretical considerations, adds an interesting dimension to the overall interpretation. This highlights the importance of a data-driven approach to cosmological studies, where feature selection can lead to more nuanced insights into dark energy models.

In the future,  we aim to extend our investigation using the cosmological constraints provided by the Dark Energy Spectroscopic Instrument (DESI).
The recent DESI Data Release 1 \citep{DESINASA} provides robust measurements of Baryon Acoustic Oscillations (BAO) in several tracers, including galaxies, quasars, and Lyman-$\alpha$ forests, over a wide redshift range from 0.1 to 4.2. These measurements provide valuable insights into the expansion history of the Universe, and place stringent constraints on cosmological parameters.
The implications of the DESI BAO measurements are of particular interest for the nature of dark energy. The DESI data, in combination with other cosmological probes such as the Planck measurements of the CMB and the Type Ia supernova datasets, may provide new perspectives on the EoS parameter of dark energy ($w$) and its possible time evolution ($w_0$ and $w_a$).
The discrepancy between the DESI BAO data and the standard $\Lambda$CDM model, especially in the context of the dark energy EoS, opens up avenues for further investigation. By incorporating the DESI BAO measurements into our analysis framework, we expect to refine our understanding of the dark energy dynamics and its implications for the overall cosmic evolution.
In addition, recent results \citep{colgain} show that a $\sim$ 2$\sigma$ discrepancy with the Planck $\Lambda$CDM cosmology in the DESI Luminous Red Galaxy (LRG) data at $z_{\text{eff}} = 0.51$ leads to an unexpectedly large $\Omega_m$ value, $\Omega_m = 0.668^{+0.180}_{-0.169}$. This anomaly causes the preference for $w_0 > -1$ in the DESI data when confronted with the $w_0w_a$CDM model. Independent analyses confirm this anomaly and show that DESI data allow $\Omega_m$ to vary on the order of $\sim$ 2$\sigma$ with increasing effective redshift in the $\Lambda$CDM model. Given the tension between LRG data at $z_{\text{eff}} = 0.51$ and Type Ia supernovae at overlapping redshifts, it is expected that this anomaly will decrease in statistical significance with future DESI data releases, although an increasing $\Omega_m$ trend with effective redshift may persist at higher redshifts. 

Recent works \citep{Alfano2024, Luongo2024, Sapone2024, Carloni2025} have tested the DESI results, in particular with respect to possible systematic biases in the BAO constraints and their compatibility with the $\Lambda$CDM model. While some analyses highlight tensions in the derived values of $\Omega_m$ and the dark energy equation of state ($w$), others argue that these discrepancies are due to systematic uncertainties rather than a fundamental departure from $\Lambda$CDM. Our approach does not aim to directly challenge or confirm these tensions, but instead provides a complementary, data-driven methodology for testing dark energy models using supernovae.

The main novelty of our work lies in the application of advanced feature selection and machine learning techniques to extract meaningful information from the supernova data sets. Rather than assuming a specific parametric form for the evolution of dark energy, our method allows for a more flexible, data-driven exploration of cosmological constraints. While our results are broadly consistent with $\Lambda$CDM, our approach provides an independent validation of existing results while highlighting the role of statistical methods in cosmology. Future incorporation of DESI data into our framework will further test whether the observed anomalies persist when analysed using our methodology.

It is worth noting that these results are not based on theoretical assumptions, but are derived directly from the data through our data-driven approach. By employing several feature selection techniques, we allow the data to guide our exploration of dark energy models. Building on these results, future research will incorporate DESI observations to further refine and develop more reliable constraints on dark energy models.

\begin{acknowledgements}
      This article is based upon work from COST Action CA21136 Addressing observational tensions in cosmology with systematic and fundamental physics (CosmoVerse) supported by COST (European Cooperation in Science and Technology). SC  acknowledges the support of {\it Istituto Nazionale di Fisica Nucleare} (INFN), {\it iniziative specifiche} MoonLight2 and QGSKY. MB acknowledges the ASI-INAF TI agreement, 2018-23-HH.0 "Attività scientifica per la missione Euclid - fase D".

      The dataset used in this study is openly accessible and can be found at \url{https://pantheonplussh0es.github.io/}. Additionally, the data is available in the public version of SNANA within the directory labeled "Pantheon+". The full SNANA dataset is archived on Zenodo and can be downloaded from \url{https://zenodo.org/record/4015325} and the SNANA source directory is \url{https://github.com/RickKessler/SNANA}.
\end{acknowledgements}

\bibliographystyle{aa} % style aa.bst
\bibliography{article} 

%-------------------------------------------------------------------

\begin{appendix} %First appendix

\onecolumn
\section{Pantheon+SH0ES features}

The total number of features provided by the dataset is 48 and are described in the Table \ref{tab:featPS}.
\begin{longtable}{|l|p{0.706\textwidth}|}
\caption{Pantheon+SH0ES features} 
\label{tab:featPS} \\
\hline
\footnotesize {Feature} & \footnotesize {Description} \\
\hline
\endfirsthead

\multicolumn{2}{c}%
{\footnotesize {\tablename\ \thetable{} -- continued from previous page}} \\
\hline
\footnotesize {Feature} & \footnotesize {Description} \\
\hline
\endhead

\hline \multicolumn{2}{r}{{\footnotesize Continued on next page}} \\
\endfoot

\hline
\endlastfoot

\footnotesize CID & \footnotesize Candidate ID \\
\footnotesize IDSURVEY & \footnotesize Survey ID \\
\footnotesize zHD & \footnotesize Hubble Diagram Redshift (with CMB and VPEC corrections) \\
\footnotesize zHDERR & \footnotesize Hubble Diagram Redshift Uncertainty \\
\footnotesize zCMB & \footnotesize CMB Corrected Redshift \\
\footnotesize zCMBERR & \footnotesize CMB Corrected Redshift Uncertainty \\
\footnotesize zHEL & \footnotesize Heliocentric Redshift \\
\footnotesize zHELERR & \footnotesize Heliocentric Redshift Uncertainty \\
\footnotesize $m_b$\_corr & \footnotesize Tripp1998 corrected/standardized $m_b$ magnitude \\
\footnotesize $m_b$\_corr\_err\_DIAG & \footnotesize $m_b$ magnitude uncertainty from the diagonal covariance matrix \\
\footnotesize MU\_SH0ES & \footnotesize Tripp1998 corrected/standardized distance modulus \\
\footnotesize MU\_SH0ES\_ERR\_DIAG & \footnotesize Uncertainty on MU\_SH0ES from the diagonal covariance matrix \\
\footnotesize CEPH\_DIST & \footnotesize Cepheid calculated absolute distance to host (incorporated in the covariance matrix) \\
\footnotesize IS\_CALIBRATOR & \footnotesize Binary indicator for SN in a host that has an associated cepheid distance \\
\footnotesize USED\_IN\_SH0ES\_HF & \footnotesize 1 if used in SH0ES 2021 Hubble Flow dataset, 0 if not included \\
\footnotesize $c$ & \footnotesize SALT2 color \\
\footnotesize $c$ERR & \footnotesize SALT2 color uncertainty \\
\footnotesize $x_1$ & \footnotesize SALT2 stretch \\
\footnotesize $x_1$ERR & \footnotesize SALT2 stretch uncertainty \\
\footnotesize $m_B$ & \footnotesize SALT2 uncorrected brightness \\
\footnotesize $m_B$ERR & \footnotesize SALT2 uncorrected brightness uncertainty \\
\footnotesize $x_0$ & \footnotesize SALT2 light curve amplitude \\
\footnotesize $x_0$ERR & \footnotesize SALT2 light curve amplitude uncertainty \\
\footnotesize COV\_$x_1\_c$ & \footnotesize SALT2 fit covariance between $x_1$ and $c$ \\
\footnotesize COV\_$x_1\_x_0$ & \footnotesize SALT2 fit covariance between $x_1$ and $x_0$ \\
\footnotesize COV\_$c\_x_0$ & \footnotesize SALT2 fit covariance between $c$ and $x_0$ \\
\footnotesize RA & \footnotesize Right Ascension \\
\footnotesize DEC & \footnotesize Declination \\
\footnotesize HOST\_RA & \footnotesize Host Galaxy RA \\
\footnotesize HOST\_DEC & \footnotesize Host Galaxy DEC \\
\footnotesize HOST\_ANGSEP & \footnotesize Angular separation between SN and host (arcsec) \\
\footnotesize VPEC & \footnotesize Peculiar velocity (km/s) \\
\footnotesize VPECERR & \footnotesize Peculiar velocity uncertainty (km/s) \\
\footnotesize MWEBV & \footnotesize Milky Way E(B-V) \\
\footnotesize HOST\_LOGMASS & \footnotesize Host Galaxy Log Stellar Mass \\
\footnotesize HOST\_LOGMASS\_ERR & \footnotesize Host Galaxy Log Stellar Mass Uncertainty \\
\footnotesize PKMJD & \footnotesize Fit Peak Date \\
\footnotesize PKMJDERR & \footnotesize Fit Peak Date Uncertainty \\
\footnotesize NDOF & \footnotesize Number of degrees of freedom in SALT2 fit \\
\footnotesize FITCHI2 & \footnotesize SALT2 fit chi squared \\
\footnotesize FITPROB & \footnotesize SNANA Fit probability \\
\footnotesize $m_b$\_corr\_err\_RAW & \footnotesize Statistical only error on fitted $m_B$ \\
\footnotesize $m_b$\_corr\_err\_VPEC & \footnotesize VPECERR propagated into magnitude error \\
\footnotesize biasCor\_$m_b$ & \footnotesize Bias correction applied to brightness $m_B$ \\
\footnotesize biasCorErr\_$m_b$ & \footnotesize Uncertainty on bias correction applied to brightness $m_B$ \\
\footnotesize biasCor\_$m_b$\_COVSCALE & \footnotesize Reduction in uncertainty due to selection effects (multiplicative) \\
\footnotesize biasCor\_$m_b$\_COVADD & \footnotesize Uncertainty floor from intrinsic scatter model (quadrature) \\
\end{longtable}
\twocolumn

\section{Additional features}

In our work we added more features to those already present in the Pantheon+SH0ES dataset in order to gain more confidence in the upcoming results. The total number of features is 74, and here we present the ones we have added.
\subsection*{Amplitude (\textit{ampl})}
The arithmetic average between the maximum and minimum magnitude:
\begin{equation}
    ampl = \frac{mag_{max} - mag_{min}}{2}.
\end{equation}
\subsection*{Beyond1std (b1std)}
The fraction of photometric points above or under a certain standard deviation from the weighted average (by photometric errors):
\begin{equation}
b1std = P(|mag - \overline{mag}| > \sigma).
\end{equation}
\subsection*{Flux Percentage Ratio (\textit{fpr})}
The percentile is the value of a variable under which there is a certain percentage of light-curve data points. The flux percentile $F_{n,m}$ was defined as the difference between the flux values at percentiles $n$ and $m$. The following flux percentile ratios have been used:
\begin{align}
    fpr20 &= F_{40,60}/F_{5,95}, \\
    fpr35 &= F_{32.5,67.5}/F_{5,95}, \\
    fpr50 &= F_{25,75}/F_{5,95}, \\
    fpr65 &= F_{17.5,82.5}/F_{5,95}, \\
    fpr80 &= F_{10,90}/F_{5,95}.
\end{align}
\subsection*{Lomb-Scargle Periodogram (\textit{ls})}
The Lomb-Scargle periodogram \citep{lomb1976least, scargle1982studies} is a method for finding periodic signals in irregularly sampled time series data. It handles irregularly spaced observations, calculates the power spectral density at different frequencies and uses least squares fitting. The statistic used in our work is the period given by the peak frequency of the Lomb-Scargle periodogram.
\subsection*{Linear Trend (\textit{slope})}
The slope of the light curve in the linear fit, that is to say the $a$ parameter in the following linear relation:
\begin{equation}
    mag = a \cdot t + b,
\end{equation} 
\begin{equation}
    slope = a.
\end{equation}
\subsection*{Median Absolute Deviation (\textit{mad})}
The median of the deviation of fluxes from the median flux:
\begin{equation}
    mad = median_i(|x_i - median_j(x_j)|).
\end{equation}
\subsection*{Median Buffer Range Percentage (\textit{mbrp})}
The fraction of data points which are within 10 per cent of the median flux:
\begin{equation}
    mbrp = P(|x_i - median_j(x_j)| < 0.1 \cdot median_j(x_j)).
\end{equation}
\subsection*{Magnitude Ratio (\textit{mr})}
An index used to estimate if the object spends most of the time above or below the median of magnitudes:
\begin{equation}
    mr = P(mag > median(mag)).
\end{equation}
\subsection*{Maximum Slope (\textit{ms})}
The maximum difference obtained measuring magnitudes at successive epochs:
\begin{equation}
    ms = max\left(\left|\frac{mag_{i+1} - mag_i}{t_{i+1} - t_i}\right|\right) = \frac{\Delta mag}{\Delta t}.
\end{equation}
\subsection*{Percent Amplitude (\textit{pa})}
The maximum percentage difference between maximum or minimum flux and the median:
\begin{equation}
    pa = max(|x_{max} - median(X)|, |x_{min} - median(X)|).
\end{equation}
\subsection*{Percent Difference Flux Percentile (\textit{pdfp})}
The difference between the second and the 98th percentile flux, converted in magnitudes. It is calculated by the ratio $F_{5,95}$ on median flux:
\begin{equation}
    pdfp = \frac{mag_{95} - mag_{5}}{median(mag)}.
\end{equation}
\subsection*{Pair Slope Trend (\textit{pst})}
The percentage of the last 30 couples of consecutive measures of fluxes that show a positive slope:
\begin{equation}
    pst = P(x_{i+1} - x_i > 0, i = n - 30,\dots, n).
\end{equation}
\subsection*{R Cor Bor (\textit{rcb})}
The fraction of magnitudes that is below 1.5 mag with respect to the median:
\begin{equation}
    rcb = P(mag > (median(mag) + 1.5)).
\end{equation}
\subsection*{Small Kurtosis (\textit{sk})}
The kurtosis represents the departure of a distribution from normality and it is given by the ratio between the fourth-order momentum and the square of the variance. For small kurtosis, it is intended the reliable kurtosis on a small number of epochs:
\begin{equation}
    sk = \frac{\mu_4}{\sigma^2}.
\end{equation}
\subsection*{Skew (\textit{skew})}
 The skewness is an index of the asymmetry of a distribution. It is given by the ratio between the third-order momentum and the variance to the third power:
 \begin{equation}
     skew = \frac{\mu_3}{\sigma^3}.
 \end{equation}
 \subsection*{Standard deviation (\textit{std})}
 The standard deviation of the fluxes.
 \subsection*{Range of a Cumulative Sum (\texorpdfstring{$R_{\text{cs}}$}{Rcs})}
 The range of a cumulative sum defined as:
 \begin{equation}
     R_{cs} = max(S) - min(S)
 \end{equation}
 \begin{equation}
     S = \frac{1}{N\sigma}\sum_{i=1}^l (mag_i - \overline{mag}).
 \end{equation}
 Where $l = 1, 2,\dots, N$.
 \subsection*{Stetson K (\textit{K})}
 A robust kurtosis measure \citep{stetson1996automatic} defined as:
 \begin{equation}
     \delta_i = \sqrt{\frac{N}{N - 1}}\frac{mag_i - \overline{mag}}{mag\_err_i},
 \end{equation}
 \begin{equation}
     K = \frac{1/N\sum_{i=1}^N|\delta_i|}{\sqrt{1/N\sum_{i=1}^N\delta_i^2}}.
 \end{equation}
 \subsection*{\texorpdfstring{${Q_{3-1}}$}{Q3-1}}
 The difference between the third and first quartile of the magnitude.
 \subsection*{Mean Variance (\textit{mvar})}
 This is a simple variability index defined as:
 \begin{equation}
     mvar = \frac{\sigma}{\overline{mag}}.
 \end{equation}
 \subsection*{CAR features (\texorpdfstring{$\sigma_C, \tau, \text{mean}$}{sigmaC, tau, mean})}
 To model irregularly sampled time series, the Continuous AutoRegressive (CAR) process, as presented in \citet{brockwell2002introduction} and \citet{falk2011first}, is employed. This continuous-time auto-regressive model involves three parameters and offers a natural and consistent means to estimate the characteristic time scale and variance of light curves. The CAR process is defined by the stochastic differential equation:
\begin{equation}
    dX(t) = -\frac{1}{\tau}X(t)dt + \sigma_C\sqrt{dt}\epsilon(t) + bdt,
\end{equation}
where ($\tau, \sigma_C, t \geq 0 $). The mean value of the light curve $X(t)$ is $b\tau$, and the variance is $\tau\sigma_C^2/2$. Here, $\tau$ is the relaxation time, interpreting the variability amplitude of the time series, and $\sigma_C$ describes variability on time scales shorter than $\tau$. $\epsilon(t)$ is a white noise process with zero mean and unit variance.

The likelihood function for such a CAR model, considering light-curve observations $\{x_1, \ldots, x_n\}$ at times $\{t_1, \ldots, t_n\}$ with measurement error variances $\{\delta_1^2, \ldots, \delta_n^2\}$, is given by:
\begin{equation}
    p(x|b, \sigma_C, \tau) = \prod_{i=1}^{n} \frac{1}{\sqrt{2\pi(\Omega_i + \delta_i^2)}} \exp\left\{-\frac{1}{2} \frac{(\hat{x}_i - x_i^*)^2}{\Omega_i + \delta_i^2}\right\},
\end{equation}
where $x^*_i = x_i - b\tau$, $\hat{x}_0 = 0$, $\Omega_0 = \tau\sigma_C^2/2$, and $x_i$ and $\Omega_i$ are given by:
\begin{equation}
    \begin{aligned}
        \hat{x}_i &= a_i \hat{x}_{i-1} + \frac{a_i\Omega_{i-1}}{\Omega_{i-1} + \delta_{i-1}^2}(x^*_{i-1} + \hat{x}_{i-1}), \\
        \Omega_i &= \Omega_0(1 - a_i^2) + a_i^2\Omega_{i-1}\left(1 - \frac{\Omega_{i-1}}{\Omega_{i-1} + \delta_{i-1}^2}\right), \\
        a_i &= e^{-(t_i - t_{i-1})/\tau}.
    \end{aligned}
\end{equation}
The parameter optimisation involves maximizing the likelihood with respect to $\sigma_C$ and $\tau$, and $b$ is determined as the mean magnitude of the light curve divided by $\tau$.

\section{Importance of XAI}

\begin{figure}
\centering
\includegraphics[width=0.91\columnwidth]{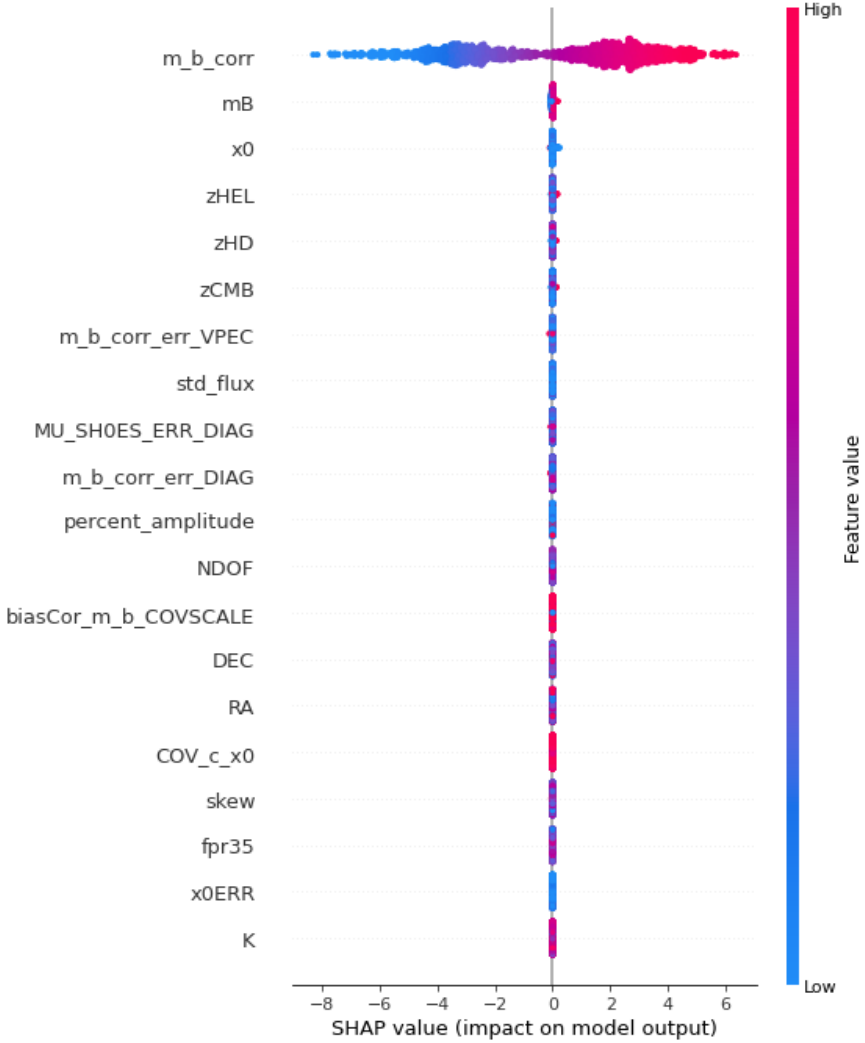}
\caption{Beeswarm plot of the 20 most important features.}
\label{fig:Beeswarm}
\end{figure}

Explainable Artificial Intelligence (XAI) plays a crucial role in improving the interpretability of machine learning models, especially in scientific applications where transparency is essential. In this appendix, we highlight the importance of XAI using the SHAP framework and present a beeswarm plot (Fig. \ref{fig:Beeswarm}) illustrating feature importances.

Explanatory AI methods such as SHAP provide insight into the decision-making process of complex models, fostering confidence and facilitating the identification of potential biases or errors. By quantifying the impact of each feature on model predictions, SHAP scores provide a comprehensive understanding of feature importance while maintaining desirable properties such as consistency and local accuracy.

To visually represent feature importance, we present a beeswarm plot showing the distribution of SHAP values for the 20 most important features. This plot provides an intuitive visualisation of the relative impact of different variables on the model's predictions, facilitating the identification of key predictors and supporting informed decision making in scientific analyses.

\section{Additional Results}

A summary of the results for the $\Lambda$CDM and CPL models is given for both the original dataset and the Boruta cases. For brevity, only the results for these specific models and cases are included in this section. The priors have been chosen on the basis of physical considerations and existing literature to ensure numerical stability while allowing meaningful parameter exploration. Some contours may appear truncated, reflecting strong constraints imposed by the data rather than artificial priors.

\begin{figure}
	\centering
	\begin{subfigure}[t]{0.4\textwidth}
		\includegraphics[width=\textwidth]{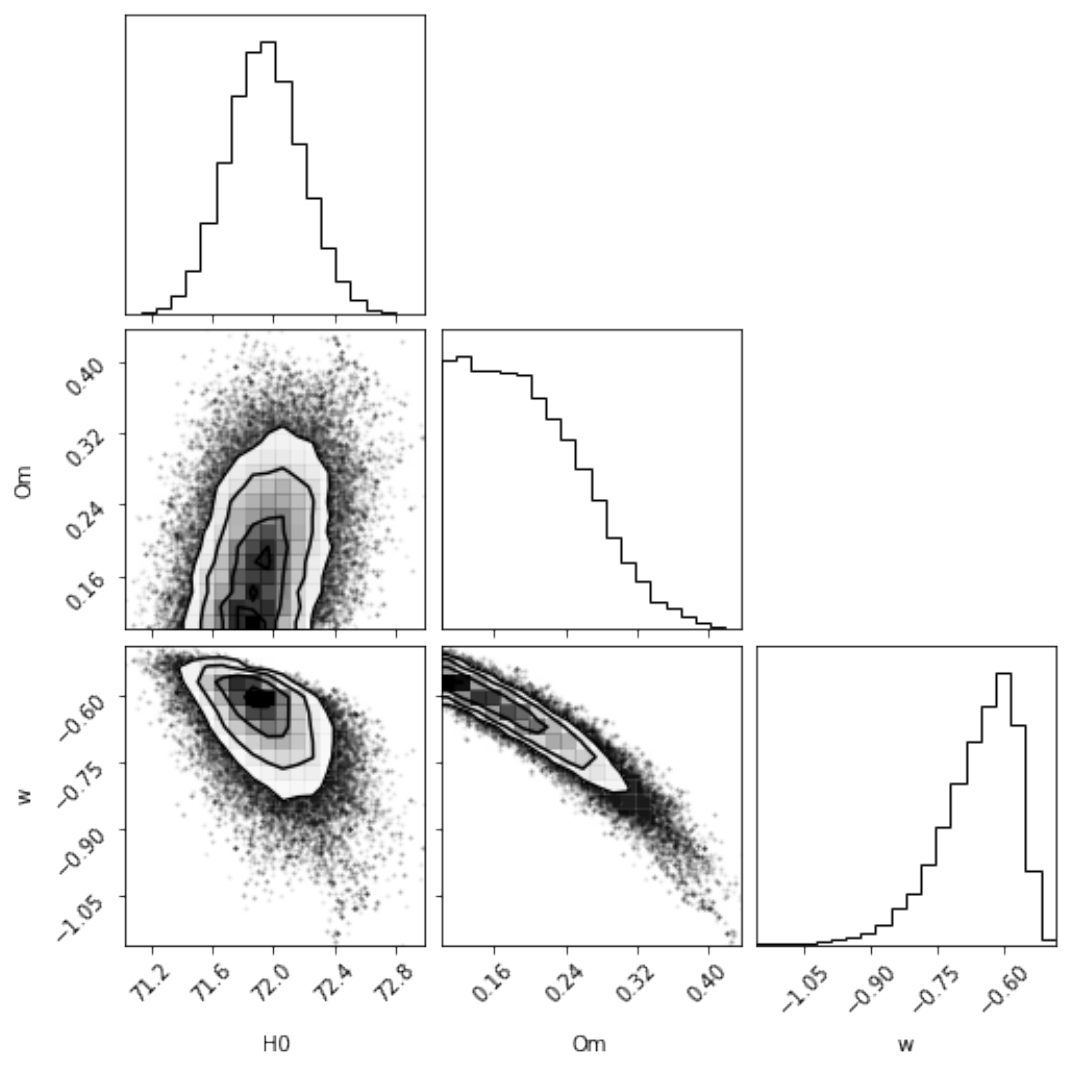}
		\caption{MCMC}
	\end{subfigure}
	\hfill
	\begin{subfigure}[t]{0.4\textwidth}
		\includegraphics[width=\textwidth]{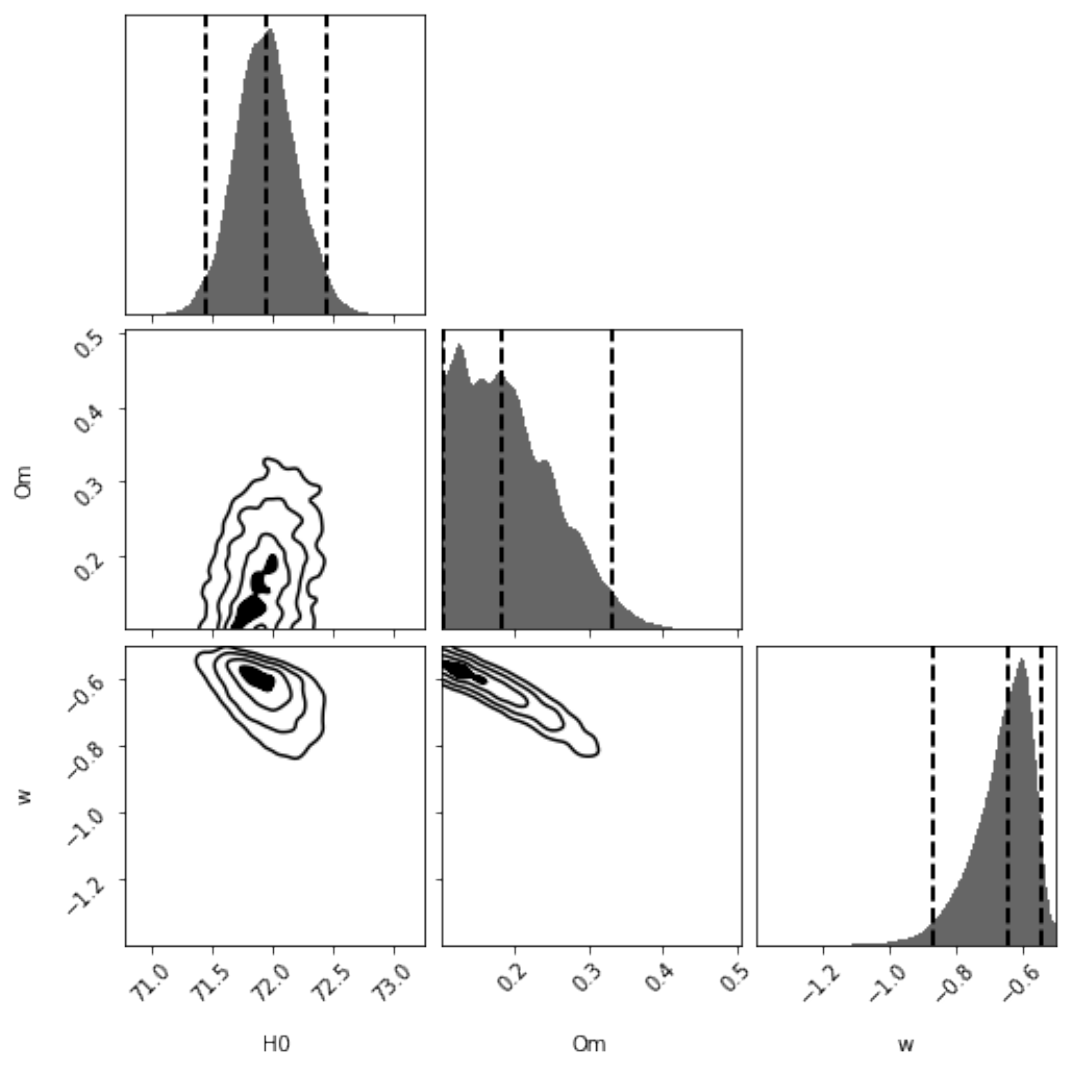}
		\caption{SNS}
	\end{subfigure}
	\hfill
	\begin{subfigure}[t]{0.4\textwidth}
		\includegraphics[width=\textwidth]{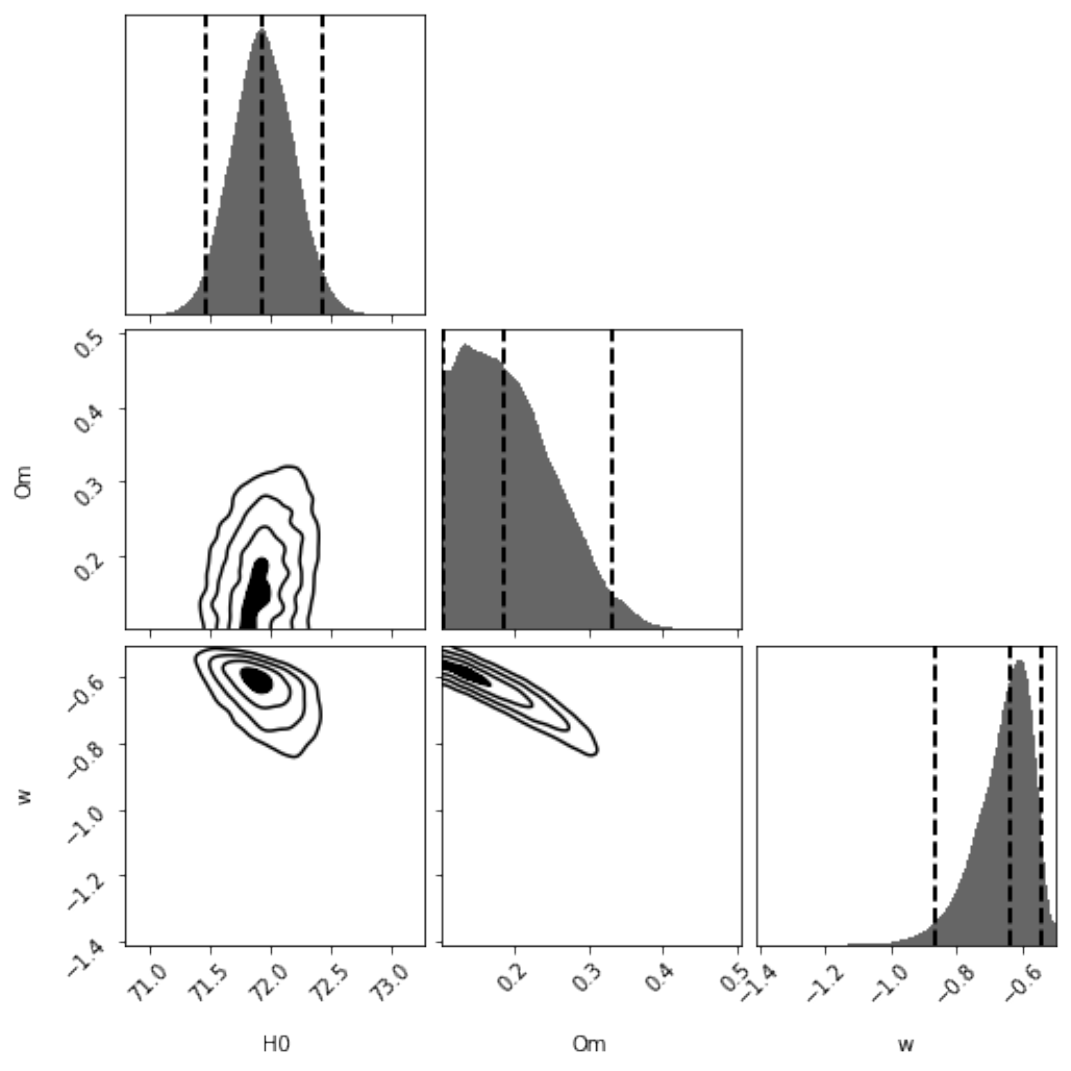}
		\caption{DNS}
	\end{subfigure}
	\caption{$\Lambda$CDM model corner plots for the original dataset.}
	\label{fig:LCDM model}
\end{figure}

\begin{figure}
	\centering
	\begin{subfigure}[t]{0.4\textwidth}
		\includegraphics[width=\columnwidth]{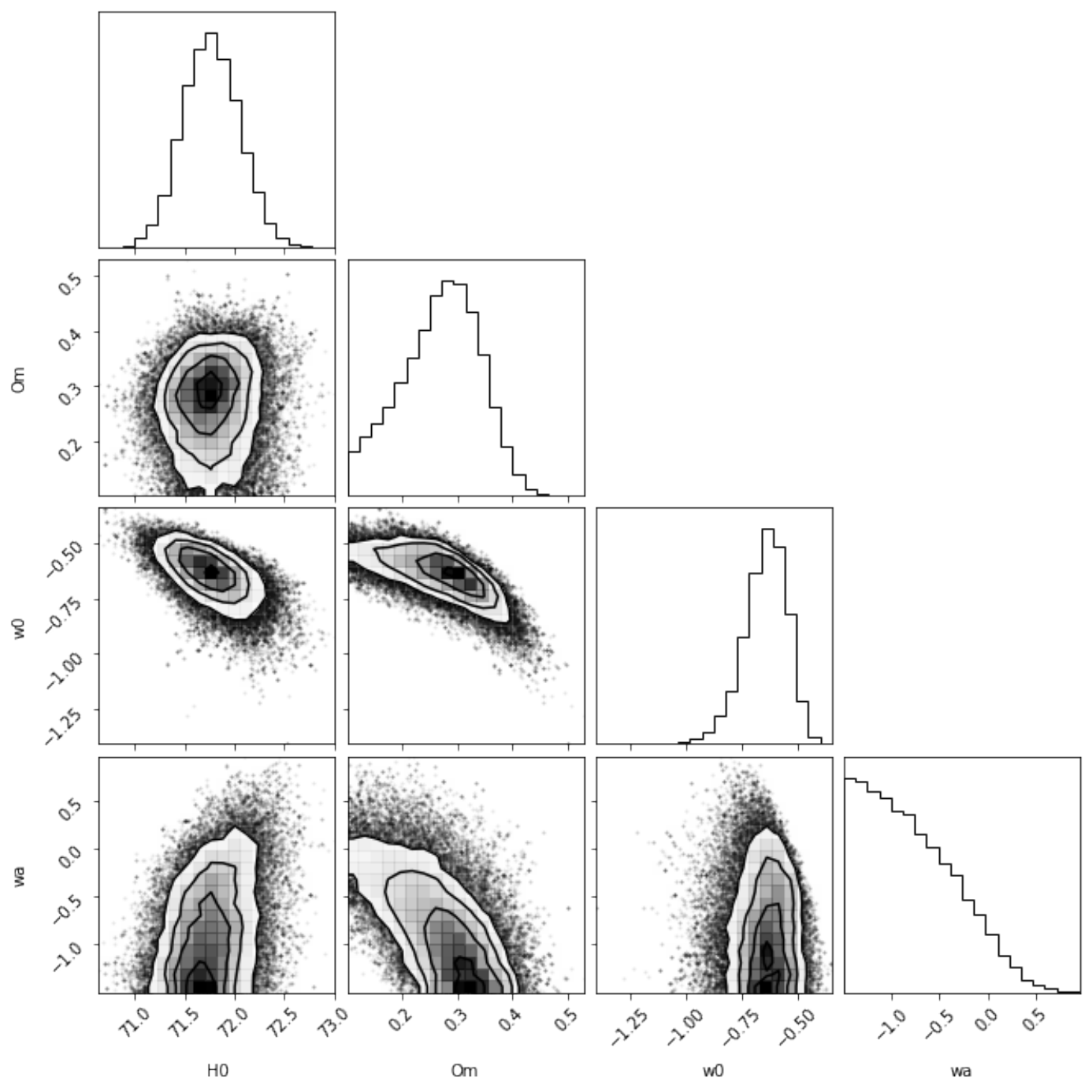}
		\caption{MCMC}
	\end{subfigure}
	\hfill
	\begin{subfigure}[t]{0.4\textwidth}
		\includegraphics[width=\textwidth]{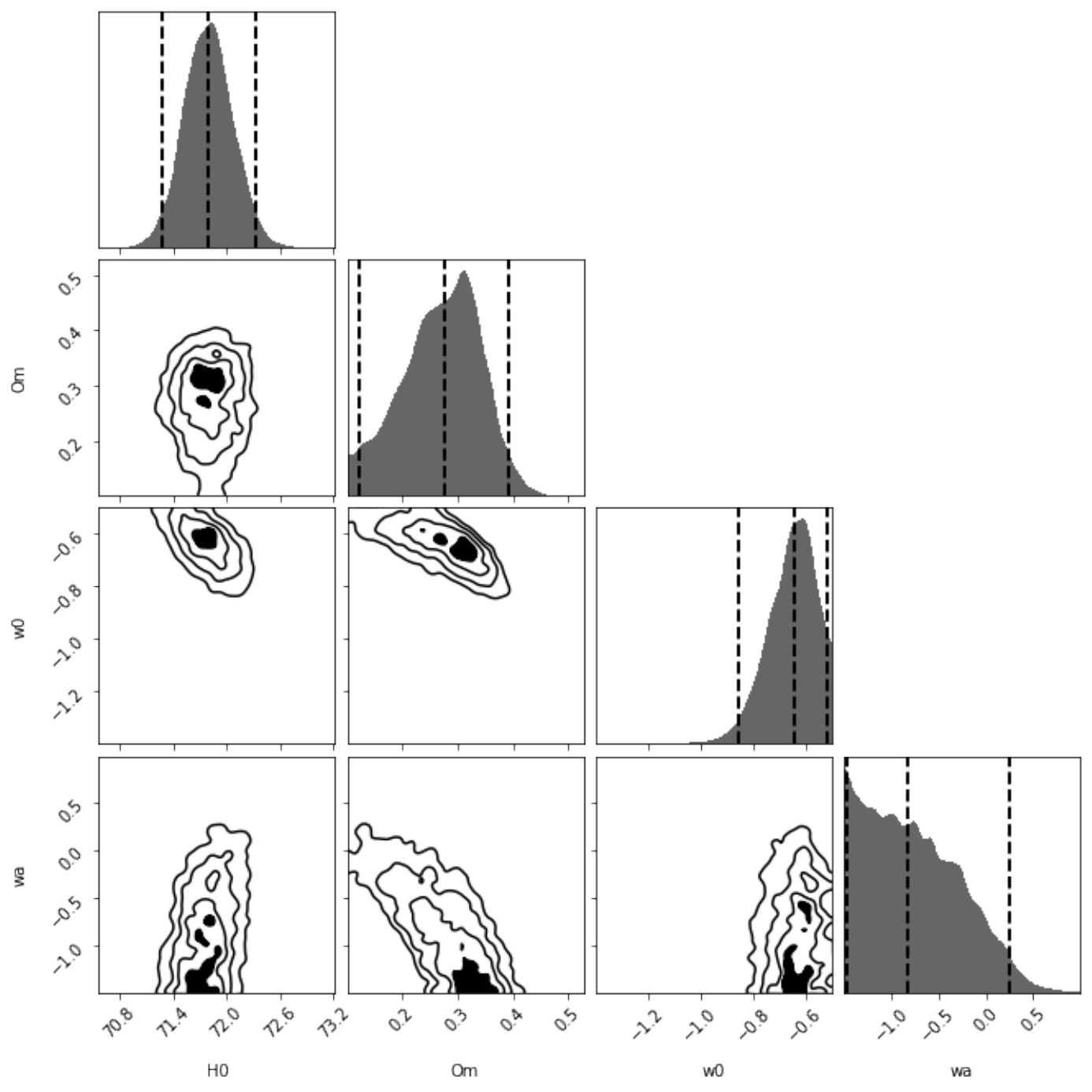}
		\caption{SNS}
	\end{subfigure}
	\hfill
	\begin{subfigure}[t]{0.4\textwidth}
		\includegraphics[width=\textwidth]{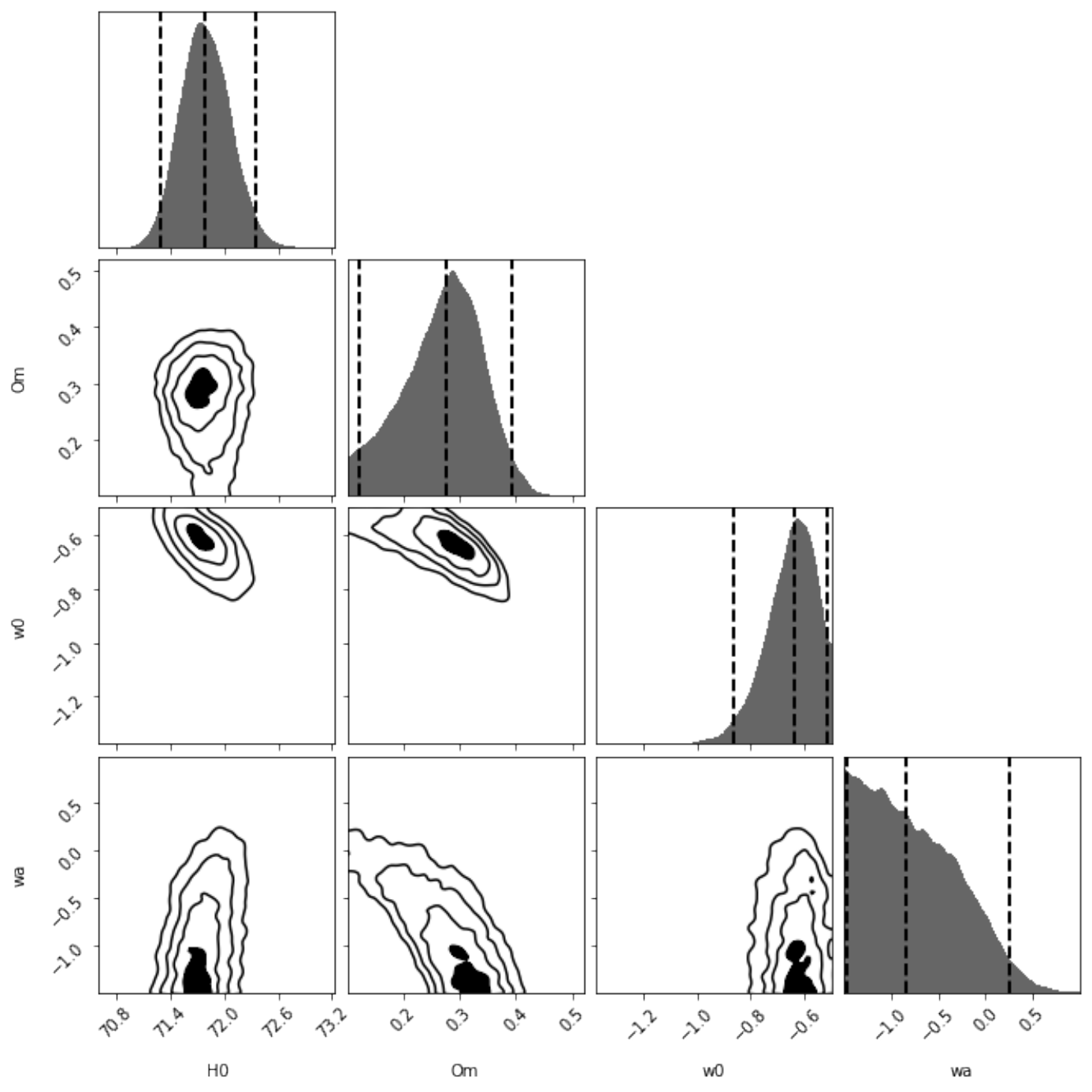}
		\caption{DNS}
	\end{subfigure}
	\caption{CPL model corner plots for the original dataset.}
	\label{fig:CPL model}
\end{figure}

\begin{figure}
	\centering
	\begin{subfigure}[t]{0.4\textwidth}
		\includegraphics[width=\textwidth]{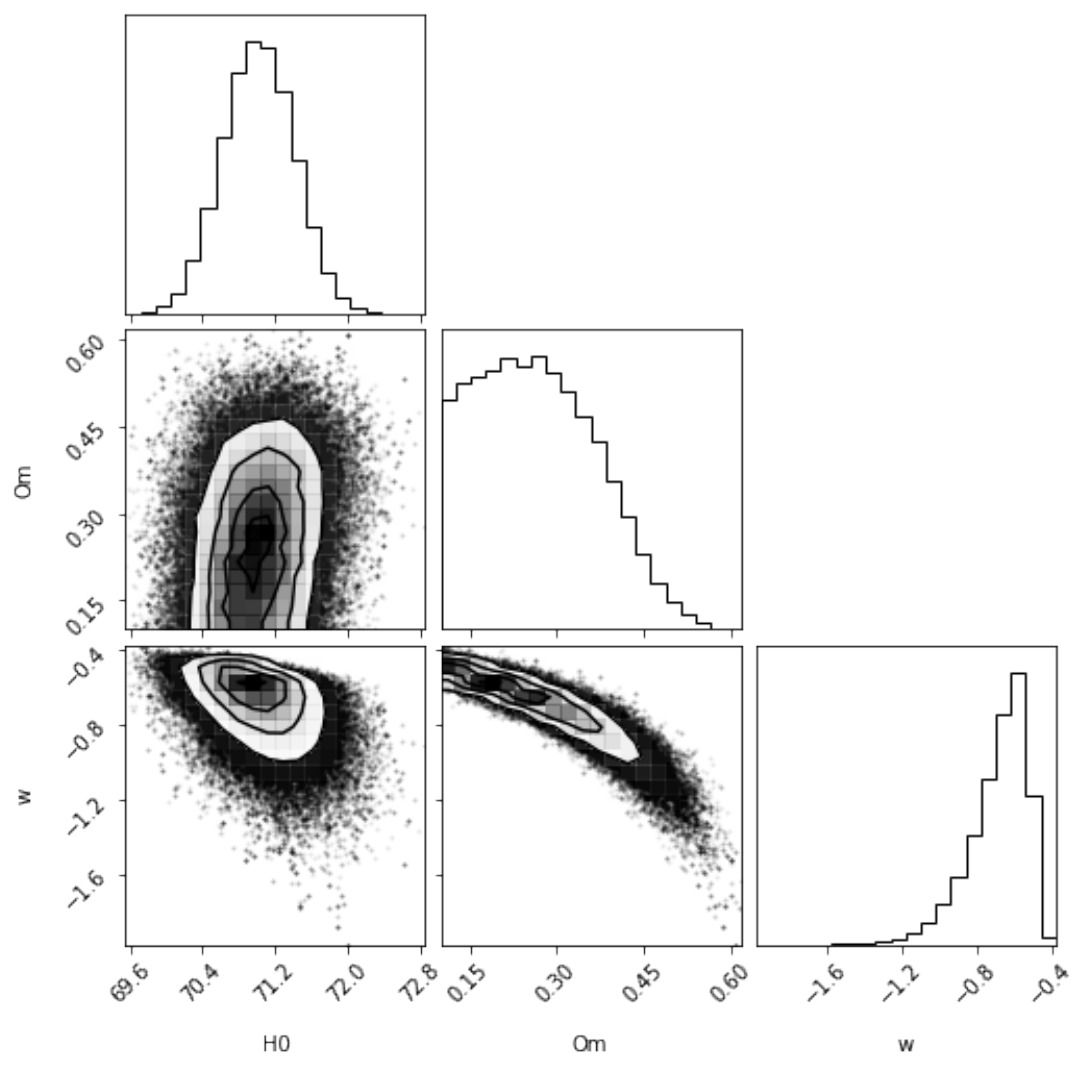}
		\caption{MCMC}
	\end{subfigure}
	\hfill
	\begin{subfigure}[t]{0.4\textwidth}
		\includegraphics[width=\textwidth]{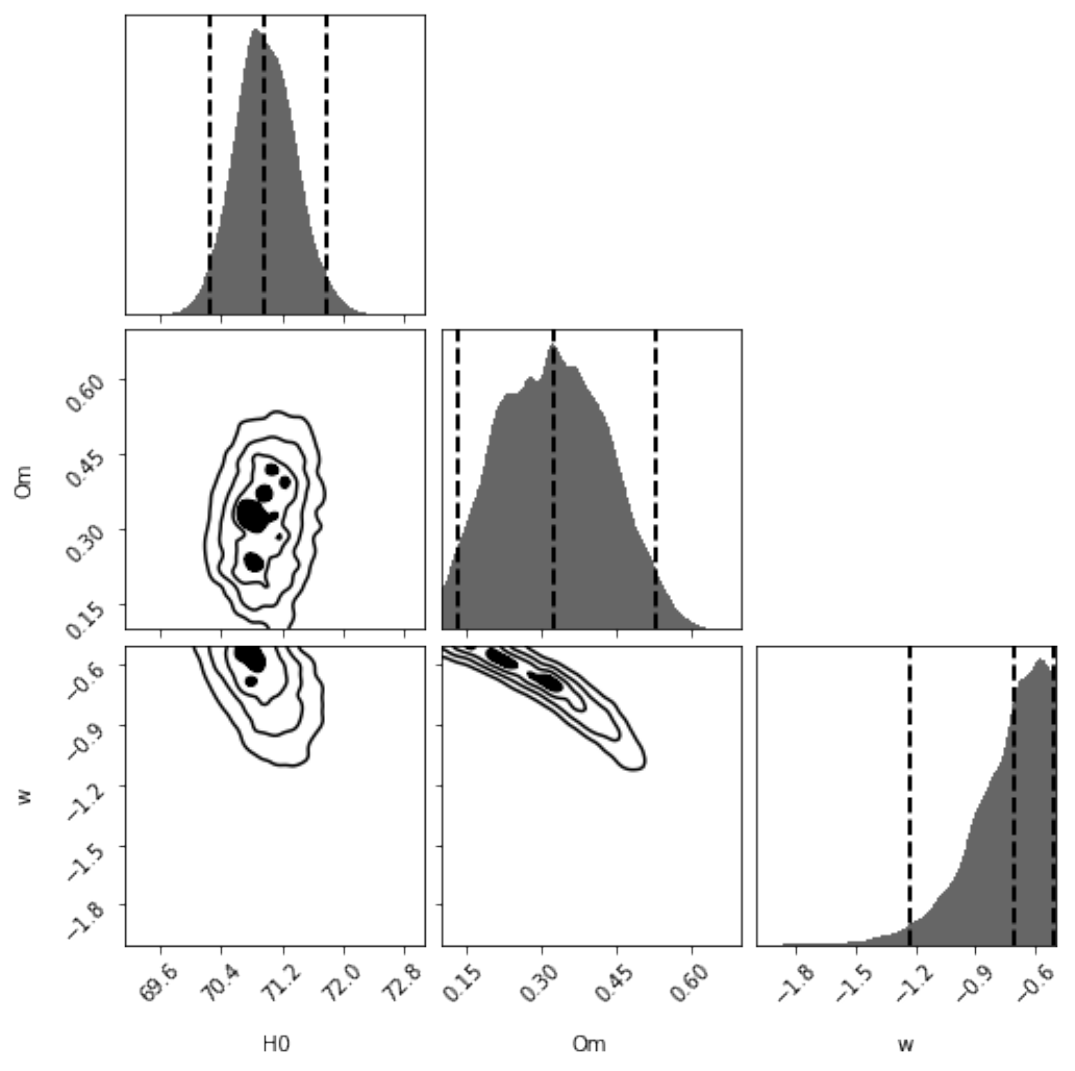}
		\caption{SNS}
	\end{subfigure}
	\hfill
	\begin{subfigure}[t]{0.4\textwidth}
		\includegraphics[width=\textwidth]{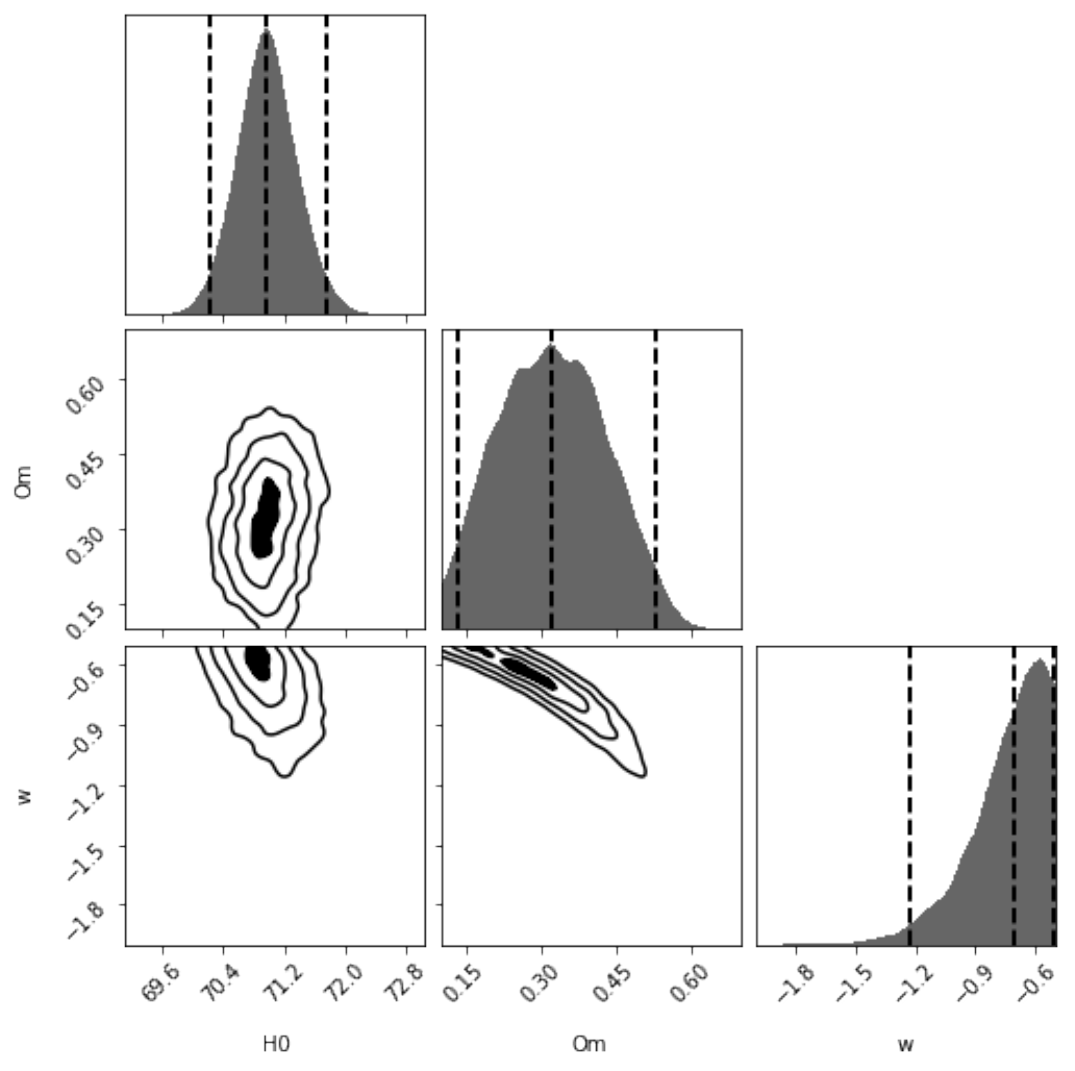}
		\caption{DNS}
	\end{subfigure}
	\caption{$\Lambda$CDM model corner plots with Boruta features.}
	\label{fig:LCDM model with Boruta features}
\end{figure}

\begin{figure}
	\centering
	\begin{subfigure}[t]{0.4\textwidth}
		\includegraphics[width=\textwidth]{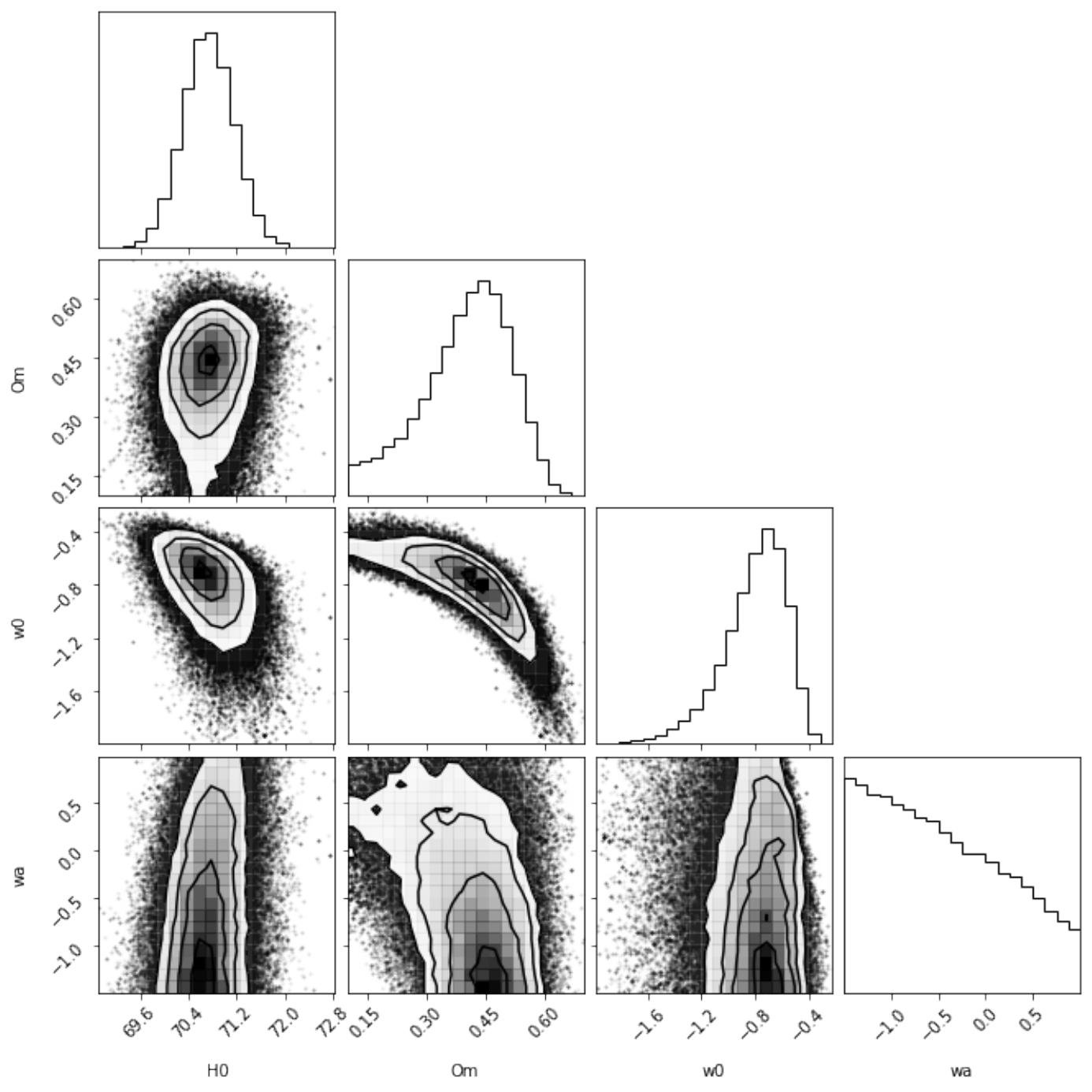}
		\caption{MCMC}
	\end{subfigure}
	\hfill
	\begin{subfigure}[t]{0.4\textwidth}
		\includegraphics[width=\textwidth]{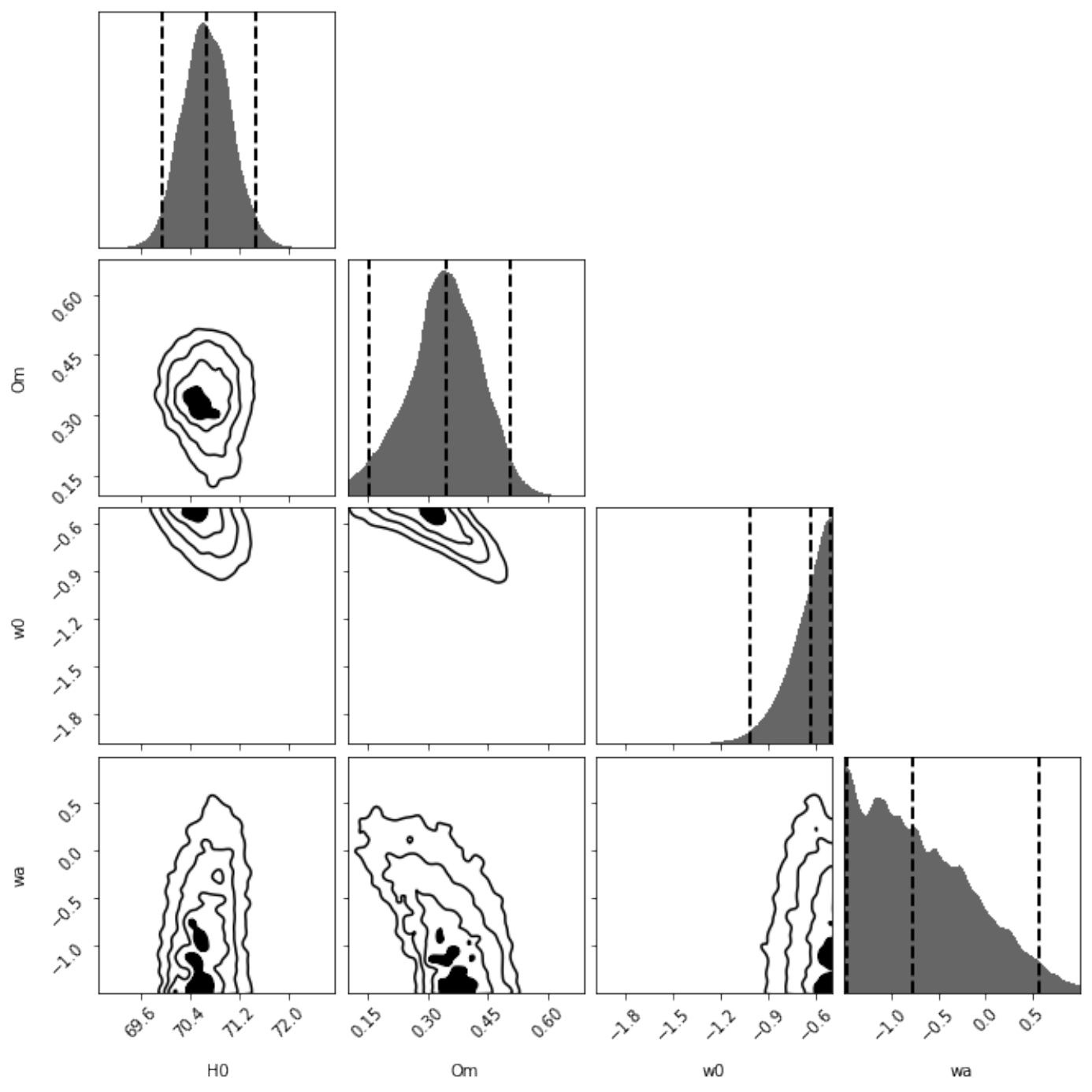}
		\caption{SNS}
	\end{subfigure}
	\hfill
	\begin{subfigure}[t]{0.4\textwidth}
		\includegraphics[width=\textwidth]{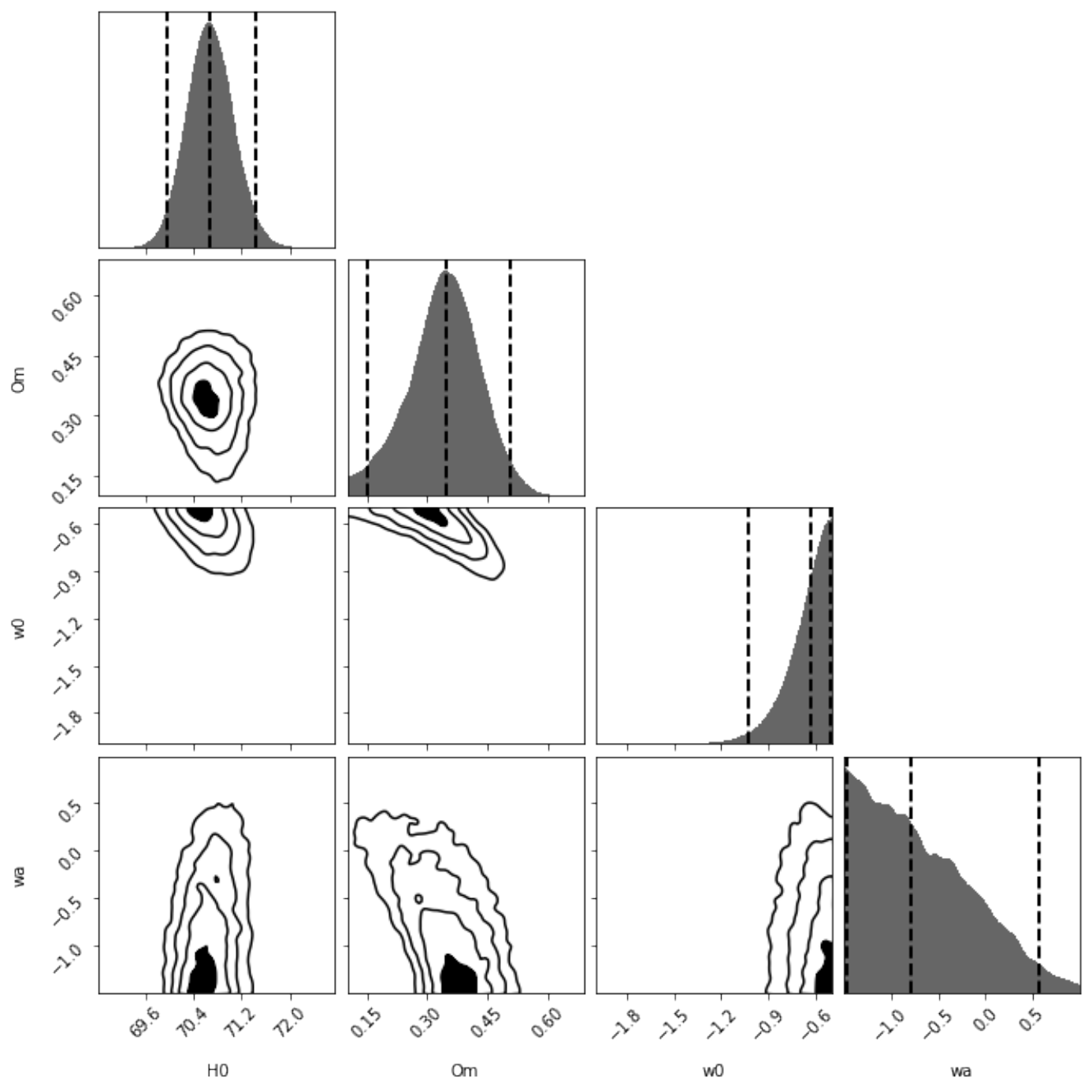}
		\caption{DNS}
	\end{subfigure}
	\caption{CPL model with Boruta features.}
	\label{fig:CPL model with Boruta features}
\end{figure}

\begin{table}
	\centering
	\begin{tabular}{lcccccc}
		\toprule
		\multirow{2}{*}{Parameter} & \multicolumn{2}{c}{MCMC} & \multicolumn{2}{c}{SNS} & \multicolumn{2}{c}{DNS} \\
		& Mean & Std & Mean & Std & Mean & Std \\
		\midrule
		$H_0$ & 71.944 & 0.249 & 71.909 & 4.701 & 71.899 & 3.409 \\
		$\Omega_m$ & 0.196 & 0.065 & 0.371 & 0.198 & 0.278 & 0.181 \\
		$w$ & -0.661 & 0.094 & -1.001 & 0.405 & -0.826 & 0.352 \\
		BIC & \multicolumn{2}{c}{2002.898} & \multicolumn{2}{c}{2002.014} & \multicolumn{2}{c}{2001.963} \\
		AIC & \multicolumn{2}{c}{1986.581} & \multicolumn{2}{c}{1985.698} & \multicolumn{2}{c}{1985.646} \\
		\midrule
		\multicolumn{7}{c}{Final Results} \\
		\midrule
		& \multicolumn{3}{c}{Mean} &\multicolumn{3}{c}{Std}\\
		\midrule
		$H_0$ & \multicolumn{3}{c}{71.917} & \multicolumn{3}{c}{1.937} \\
		$\Omega_m$ & \multicolumn{3}{c}{0.281} & \multicolumn{3}{c}{0.092} \\
		$w$ & \multicolumn{3}{c}{-0.829} & \multicolumn{3}{c}{0.182} \\
		\bottomrule
	\end{tabular}
	\caption{$\Lambda$CDM model parameter values for the original dataset.}
\end{table}

\begin{table}
	\centering
	\begin{tabular}{lcccccc}
		\toprule
		\multirow{2}{*}{Parameter} & \multicolumn{2}{c}{MCMC} & \multicolumn{2}{c}{SNS} & \multicolumn{2}{c}{DNS} \\
		& Mean & Std & Mean & Std & Mean & Std \\
		\midrule
		$H_0$ & 71.774 & 0.289 & 71.874 & 4.538 & 71.807 & 3.329 \\
		$\Omega_m$ & 0.264 & 0.071 & 0.390 & 0.183 & 0.338 & 0.150 \\
		$w_0$ & -0.643 & 0.099 & -0.985 & 0.405 & -0.825 & 0.346 \\
		$w_a$ & -0.781 & 0.489 & -0.405 & 0.687 & -0.656 & 0.649 \\
		BIC & \multicolumn{2}{c}{2008.777} & \multicolumn{2}{c}{2007.535} & \multicolumn{2}{c}{2007.470} \\
		AIC & \multicolumn{2}{c}{1987.021} & \multicolumn{2}{c}{1985.779} & \multicolumn{2}{c}{1985.714} \\
		\midrule
		\multicolumn{7}{c}{Final Results} \\
		\midrule
		& \multicolumn{3}{c}{Mean} &\multicolumn{3}{c}{Std}\\
		\midrule
		$H_0$ & \multicolumn{3}{c}{71.818} & \multicolumn{3}{c}{1.879} \\
		$\Omega_m$ & \multicolumn{3}{c}{0.331} & \multicolumn{3}{c}{0.082} \\
		$w_0$ & \multicolumn{3}{c}{-0.817} & \multicolumn{3}{c}{0.181} \\
		$w_a$ & \multicolumn{3}{c}{-0.614} & \multicolumn{3}{c}{0.355} \\
		\bottomrule
	\end{tabular}
	\caption{CPL model parameter values for the original dataset.}
\end{table}

\begin{table}
	\centering
	\begin{tabular}{lcccccc}
		\toprule
		\multirow{2}{*}{Parameter} & \multicolumn{2}{c}{MCMC} & \multicolumn{2}{c}{SNS} & \multicolumn{2}{c}{DNS} \\
		& Mean & Std & Mean & Std & Mean & Std \\
		\midrule
		$H_0$ & 70.957 & 0.392 & 71.130 & 5.301 & 71.039 & 4.116 \\
		$\Omega_m$ & 0.284 & 0.107 & 0.456 & 0.192 & 0.359 & 0.191 \\
		$w$ & -0.694 & 0.174 & -1.088 & 0.417 & -0.890 & 0.406 \\
		BIC & \multicolumn{2}{c}{352.940} & \multicolumn{2}{c}{351.903} & \multicolumn{2}{c}{351.897} \\
		AIC & \multicolumn{2}{c}{341.444} & \multicolumn{2}{c}{340.408} & \multicolumn{2}{c}{340.401} \\
		\midrule
		\multicolumn{7}{c}{Final Results} \\
		\midrule
		& \multicolumn{3}{c}{Mean} &\multicolumn{3}{c}{Std}\\
		\midrule
		$H_0$ & \multicolumn{3}{c}{71.042} & \multicolumn{3}{c}{3.270} \\
		$\Omega_m$ & \multicolumn{3}{c}{0.366} & \multicolumn{3}{c}{0.163} \\
		$w$ & \multicolumn{3}{c}{-0.890} & \multicolumn{3}{c}{0.332} \\
		\bottomrule
	\end{tabular}
	\caption{$\Lambda$CDM parameter values with Boruta features.}
\end{table}

\begin{table}
	\centering
	\begin{tabular}{lcccccc}
		\toprule
		\multirow{2}{*}{Parameter} & \multicolumn{2}{c}{MCMC} & \multicolumn{2}{c}{SNS} & \multicolumn{2}{c}{DNS} \\
		& Mean & Std & Mean & Std & Mean & Std \\
		\midrule
		$H_0$ & 70.848 & 0.424 & 71.105 & 5.210 & 70.966 & 3.910 \\
		$\Omega_m$ & 0.331 & 0.101 & 0.459 & 0.185 & 0.414 & 0.154 \\
		$w_0$ & -0.672 & 0.185 & -1.067 & 0.415 & -0.886 & 0.389 \\
		$w_a$ & -0.693 & 0.585 & -0.303 & 0.711 & -0.635 & 0.720 \\
		BIC & \multicolumn{2}{c}{357.482} & \multicolumn{2}{c}{355.805} & \multicolumn{2}{c}{355.618} \\
		AIC & \multicolumn{2}{c}{342.154} & \multicolumn{2}{c}{340.478} & \multicolumn{2}{c}{340.291} \\
		\midrule
		\multicolumn{7}{c}{Final Results} \\
		\midrule
		& \multicolumn{3}{c}{Mean} &\multicolumn{3}{c}{Std}\\
		\midrule
		$H_0$ & \multicolumn{3}{c}{70.973} & \multicolumn{3}{c}{3.181} \\
		$\Omega_m$ & \multicolumn{3}{c}{0.401} & \multicolumn{3}{c}{0.147} \\
		$w_0$ & \multicolumn{3}{c}{-0.875} & \multicolumn{3}{c}{0.329} \\
		$w_a$ & \multicolumn{3}{c}{-0.544} & \multicolumn{3}{c}{0.672} \\
		\bottomrule
	\end{tabular}
	\caption{CPL parameter values with Boruta features.}
\end{table}
\end{appendix}
\end{document}